\documentclass[aps,pre, 11pt,reprint,onecolumn, nobibnotes,longbibliography,superscriptaddress]{revtex4-2} 
\usepackage{amsmath,amssymb,amsfonts,dsfont,bm} 
\usepackage{physics} 
\usepackage{graphicx,xcolor} 
\usepackage[colorlinks=true,linkcolor=blue,citecolor=blue,urlcolor=blue]{hyperref}
\usepackage[all]{hypcap} 

\begin{document} 
\title{Local quantum coherence with intersource interactions at nonzero temperature} 

\author{Yehor Hudenko}
\email{yehhuden@gmail.com} 
\affiliation{Charles University, Faculty of Mathematics and Physics, Department of Macromolecular Physics, V Hole\v{s}ovi\v{c}k\'{a}ch~2, CZ-18000 Praha 8, Czech Republic}

\author{Michal Kol\'{a}\v{r}}
\email{kolar@optics.upol.cz} 
\affiliation{Palack\'{y} University, Department of Optics, 17.~listopadu 1192/12, 771~46~Olomouc, Czech Republic}

\author{Radim Filip}
\email{filip@optics.upol.cz} 
\affiliation{Palack\'{y} University, Department of Optics, 17.~listopadu 1192/12, 771~46~Olomouc, Czech Republic}

\author{Artem Ryabov}
\email{artem.ryabov@matfyz.cuni.cz} 
\affiliation{Charles University, Faculty of Mathematics and Physics, Department of Macromolecular Physics, V Hole\v{s}ovi\v{c}k\'{a}ch~2, CZ-18000 Praha 8, Czech Republic}

\date{November~20, 2026} 

\begin{abstract} 
Local quantum coherence in a two-level system (TLS) is typically generated via time-dependent driving. However, it can also emerge autonomously from symmetry-breaking interactions between the TLS and its surrounding environment at a low temperature. Although such environments often consist of interacting atoms or spins, the role of interactions within the environment in generating the autonomous local coherence has remained unexplored. Here, we address this gap by analyzing an exactly solvable model, which comprises a target TLS coupled to $N$ interacting source TLSs that represent the environment, with the whole system being in thermal equilibrium. We show that the local coherence not only persists but can be enhanced at finite temperatures of the environment compared to the case of no intersource interactions. The temperature dependence of the coherence bears signatures of a ground-state level crossing, and our analytical results suggest strategies for its optimization. Our findings reveal generic properties of the autonomously generated quantum coherence and point to viable routes for observing the coherence at nonzero temperatures. 
\end{abstract} 
\maketitle 

\section{Introduction}
Quantum coherences~\cite{Streltsov/etal:2017, Chitambar/Gilad:2019} represent essential resources in modern quantum technologies~\cite{Boixo/etal:2018, Arute/etal:2019, Arisoy/etal:2021, Solfanelli/etal:2021, Hangleiter/Eisert:2023, Ullah/etal:2023} that are amenable to experimental manipulation in multiple physical systems~\cite{Hofheinz/etal:2009, Regula/etal:2018, Fang/etal:2018, Wu/etal:2018, Gumberidze/etal:2019, Starek/etal:2021, Maillette/etal:2025}. They may also play an important role in quantum thermodynamics~\cite{Lostaglio/etal:2015, Narasimhachar/Gilad:2015, Klatzow/etal:2019, Hammam/etal:2022, Aamir/etal:2025}, quantum biology~\cite{Huelga/etal:2013, Cao/etal:2020}, and quantum  gravity~\cite{Hosten:2022}. Generation of the coherences in a target system is typically achieved through highly intense external drives, such as lasers or masers. These nonautonomous methods often come with high energy or power costs and require precise time-resolved external control of quantum states. 

Hence, it would be advantageous to develop autonomous mechanisms for generating quantum coherences not relying on the continuous external driving. In Refs.~\cite{Guarnieri/etal:2018, Guarnieri/etal:2020, Purkayastha/etal:2020, Cresser/Anders:2021, Slobodeniuk/etal:2021, Slobodeniuk/etal:2024, Hogg/etal:2024}, it was suggested to address this goal by engineering suitable strong interactions between a target two-level system (TLS) and a reservoir composed of an infinite number of bosonic modes or of TLSs~\cite{Roman-Ancheyta/etal:2021}. These studies identified classes of system-reservoir interactions capable of inducing stationary coherences in the energy eigenbasis of the target TLS, with the coherences being largest at low reservoir temperatures. To date, no experimental demonstration of this mechanism for autonomous steady-state coherence generation has been reported, likely due to the difficulty of implementing the required TLS-reservoir couplings.

An alternative approach to realizing the steady-state coherence generation involves coupling the target TLS to a finite number $N$ of auxiliary quantum systems that act as sources of the coherence. These source systems can be regarded as part of the immediate environment of the target TLS. A representative model of this type has been introduced in Ref.~\cite{Kolar/Filip:2024}, where the sources were modeled as TLSs with the energy gap $\hbar \omega_{\rm a}$. In this scheme, the $i$th source TLS (described by spin operators $\hat\sigma_i^{x,y,z}$) interacts with the target TLS ($\hat\sigma_0^{x,y,z}$) through the symmetry-breaking interaction proportional to $\hat\sigma_0^x \otimes \hat\sigma_i^z$, while the source TLSs do not interact among themselves. The composite target–source system is assumed to be in a thermal equilibrium state with a weakly-coupled environment at temperature $T$. The main conclusion of Ref.~\cite{Kolar/Filip:2024} is that the parameters $N$ and $\omega_{\rm a}$ are central for generating nonzero steady-state coherence at finite temperatures. Increasing either parameter leads to stronger steady-state coherence at $T>0$, hence improving the prospects for observing autonomous coherence generation under realistic thermal conditions.

However, the model of Ref.~\cite{Kolar/Filip:2024} was studied under the simplifying assumption of noninteracting source TLSs. This leaves unresolved the crucial question: To what extent do the intersource (intra-environment) interactions, naturally occurring in realistic condensed-matter systems, affect the steady-state coherence in the target TLS? The present work addresses this issue by systematically analyzing various effects arising due to the mutual interaction of the source systems. 

As a minimal yet nontrivial model for the intra-environment interactions, we consider the nearest-neighbor spin-spin coupling, corresponding to the Ising-type interaction~\cite{Stinchcombe:1973, Samaj:2010, Campostrini/etal:2015}. Even within this basic framework, the coherence generated in the target TLS exhibits a remarkably rich behavior, qualitatively distinct from that found in the noninteracting-source scenario~\cite{Kolar/Filip:2024} or in the reservoir-based models~\cite{Guarnieri/etal:2018, Guarnieri/etal:2020, Purkayastha/etal:2020, Cresser/Anders:2021, Slobodeniuk/etal:2021, Slobodeniuk/etal:2024, Hogg/etal:2024, Roman-Ancheyta/etal:2021}. In particular, ground-state energy-level crossing in the composite system gives rise to nonmonotonic temperature dependence, where the coherence can increase with temperature, reaching a maximum at finite $T>0$. Moreover, the interaction can amplify the coherence substantially at moderate temperatures.

\section{Model and quantities of interest}

We consider a target two-level system, characterized by spin operators $\hat\sigma_0^{x,y,z}$ and an energy gap $\omega_0$ ($\hbar\! =\! 1$ in the following), interacting with a chain of $N$ source two-level systems. The $i$th source TLS has spin operators $\hat\sigma_i^{x,y,z}$ and an energy gap $\omega_{\rm a}$, $i = 1, \ldots, N$. The total Hamiltonian is given by:
\begin{align}  
\label{eq:Htot} 
\hat H= \frac{\omega_0}{2} \hat\sigma_0^z 
+ \frac{\omega_{\rm a}}{2} \sum_{i=1}^N  \hat\sigma_i^z 
+\frac{\gamma}{2} \sum_{i=1}^N \hat\sigma_0^x \hat\sigma_i^z  
+\hat H^{\rm{aa}}_{\rm int}, 
\end{align} 
where $\gamma$ is the coupling constant of interaction between the target and the source TLSs~\cite{Kolar/Filip:2024}, and 
\begin{equation} 
\label{eq:Hint-Ising}
\hat{H}^{\rm aa}_{\rm int}=-\frac{J}{2}\sum_{i=1}^N\hat\sigma_i^z \hat\sigma_{i+1}^z 
\end{equation}
represents the nearest-neighbor intersource interaction of the Ising type~\cite{Samaj:2010} subject to periodic boundary conditions $\hat{\sigma}^{z}_{N+1}=\hat{\sigma}^{z}_1$. 

We assume that the composite system with the Hamiltonian~\eqref{eq:Htot} resides in the thermodynamic equilibrium described by the canonical density operator  
\begin{equation}
\label{eq:def_totalrho}
\hat\rho = \frac{\exp(-\beta \hat H)}{{\rm tr}\!\left[\exp(-\beta \hat H)\right]},
\end{equation}
with $\beta = 1/k_{\rm B}T$ denoting the inverse thermal energy. In such a state, the reduced density operator $\hat\rho_0$ of the target TLS is calculated as the partial trace 
\begin{equation} 
\label{eq:rho0-def}
\hat\rho_0 =\displaystyle {\rm tr}_{\rm a} (\hat\rho)  
\end{equation} 
over the Hilbert space of the source TLSs. We represent $\hat\rho_0$ by a matrix in the basis $\{\ket{e},\ket{g}\}$ of eigenvectors of the spin operator $\hat\sigma_0^z$:   
$\hat\sigma_0^z \ket{e} = +1 \ket{e}$, 
$\hat\sigma_0^z \ket{g} = -1 \ket{g}$, i.e., 
corresponding to the excited and ground state of a free target TLS: 
\begin{equation}
\label{eq:rho0-def-matrix}
\hat{\rho}_{0}=\begin{pmatrix}
\rho_e & \rho_{ge} \\ \rho_{eg} & \rho_g
\end{pmatrix}, 
\end{equation} 
with $\rho_e$ ($\rho_g$) being the population of state $\ket{e}$ ($\ket{g}$). 

According to Eq.~\eqref{eq:def_totalrho}, the composite system consisting of the target TLS and $N$ sources is in the global canonical Gibbs state at temperature $T$. Physically, this state arises when the composite system is weakly coupled to an equilibrium super-reservoir, such as a phonon bath, and is attained on timescales longer than all relaxation times of the composite system. The relaxation dynamics toward the global equilibrium may be described by a standard Born-Markov master equation, in which the system and reservoir states are initially factorized, and the reservoir degrees of freedom are traced out~\cite{Trushechkin/etal:2022}. In equilibrium, the composite system has no coherences in its energy eigenbasis, whereas the reduced state of the target TLS may retain autonomous steady-state quantum coherence without external time-dependent driving or additional energy input.

Our main quantity of interest will be the quantum coherence in the target TLS in the basis $\{\ket{e}, \ket{g}\}$. As its measure $C$, we take 
\begin{equation} 
\label{eq:C-definition}
C=|\rho_{ge}|+|\rho_{eg}|, 
\end{equation} 
with $0\leq C\leq 1$. We will discuss the behavior of $C$ as a function of $T$, $J$, $\omega_{\rm a}$, $\gamma$, and $N$; the target TLS gap $\omega_0$ will be used to define the energy unit. 

The central questions addressed in this work are how interactions among multiple source spins can generate, enhance, and protect steady-state local coherence in the target TLS at finite temperature and in the absence of externally imposed coherence. Several related but distinct problems have been studied for Hamiltonians sharing some qualitative features with $\hat H$ in Eq.~\eqref{eq:Htot}: 
The time-dependent loss of coherence of a target spin coupled to a spin environment at $T=0$ (unitary evolution) has been investigated in Refs.~\cite{Cucchietti/etal:2005, Quan/etal:2006, Cucchietti/etal:2007, Rossini/etal:2007, Cormick/Paz:2008}, see also references therein. Refs.~\cite{Cucchietti/etal:2005, Hutton/Bose:2004, Breuer/etal:2004} consider a noninteracting spin environment, which allows for analytical treatment of the system dynamics, including derivation of the exact reduced dynamics of a central spin~\cite{Breuer/etal:2004}. Ref.~\cite{Quan/etal:2006} studies a transverse-field Ising spin chain and finds that decoherence is sharply enhanced near its quantum critical point. Ref.~\cite{Cucchietti/etal:2007} explores the decoherence when the target-environment coupling drives the spin environment through a quantum phase transition. 
Ref.~\cite{Rossini/etal:2007} investigates decoherence of a qubit coupled to Ising, XY, and Heisenberg chains, including cases in which the qubit interacts with a part of the chain only, also focusing on how decoherence changes as the bath enters different phases. Ref.~\cite{Cormick/Paz:2008} focuses on the XY-type interactions and shows that a coupling-independent decoherence regime is not, by itself, evidence of a quantum phase transition, since such a regime can also occur in its absence; it further finds that, at strong coupling, the transition does not enhance decoherence.

Note that in the present Hamiltonian~\eqref{eq:Htot}, the target-source coupling $\sim\hat\sigma_0^x\hat\sigma_i^z$ does not commute with the target self-energy $\omega_0\hat\sigma_0^z/2$. Consequently, tracing out the source spins leaves nonzero coherence in the target, including in the finite-temperature equilibrium state. As shown in Sec.~\ref{sec:C-antiferromagnetic-even}, and consistent with the findings of Ref.~\cite{Cucchietti/etal:2007}, the presence of the target TLS can shift the transition between the ferromagnetic and antiferromagnetic ground states. Since the spin environment is classical in our model, this transition is a level-crossing transition, not a quantum critical point.

\section{Eigenvalues and eigenvectors of the total Hamiltonian} 

To solve the eigenvalue problem 
\begin{equation}
    \hat{H}\ket\tau = E\ket\tau 
\end{equation}
for the total Hamiltonian $\hat{H}$ given in Eq.~\eqref{eq:Htot}, we introduce the operator 
\begin{equation}{\label{eq:Sa-def}}
\hat S_{\rm a} = \sum_{i=1}^N \hat\sigma_i^z  
\end{equation} 
representing the total magnetization of source TLSs. Its spectrum is degenerate with eigenvalues $2s$ and eigenvectors $\ket{s,m}$, 
\begin{equation}
\label{eq:Sa-eigenproblem}
\hat{S}_{\rm a}\ket{s,m}=2s \ket{s,m},\quad s\in\{-N/2,-N/2+1,\ldots ,N/2\}, 
\end{equation} 
where $m$ labels distinct eigenstates corresponding to the same value of $s$. The state $\ket{s,m}$ can be expressed through states $\ket{\alpha_i}$ of $i$-th source TLS, as
\begin{equation}
\ket{s,m}=\ket{\alpha_1} \ldots \ket{\alpha_N},
\end{equation} 
where $\ket{\alpha_i}\in\{\ket{e},\ket{g}\}$ are the eigenstates of $\hat{\sigma}_i^z$. 

The total Hamiltonian~\eqref{eq:Htot} then assumes the form
\begin{align}  
\label{eq:Htot-Ising-S} 
\hat H= \frac{\omega_0}{2} \hat\sigma_0^z 
+ \frac{\omega_{\rm a}}{2} \hat S_{\rm a}
+\frac{\gamma}{2}\hat\sigma_0^x \hat S_{\rm a} 
-\frac{J}{2}\sum_{i=1}^N\hat\sigma_i^z \hat\sigma_{i+1}^z.
\end{align}  
In App.~\ref{sec:appendixA} we show that eigenvectors $\ket\tau$ of this Hamiltonian may be factorized:   
\begin{equation}
\label{eq:tau-factorized}
\ket\tau = \ket\alpha \ket{\psi}, 
\end{equation} 
where  
\begin{equation}
\label{eq:alpha-definition}
\ket{\alpha}=c_1\ket{e}+c_2\ket{g},\quad 
|c_1|^2+|c_2|^2=1,
\end{equation} 
is the state of the target TLS with complex weighting factors $c_1$ and $c_2$, and $\ket{\psi}$ is the state of source TLSs. 
Moreover, we have $\ket{\psi}=\ket{s,m}$, due to the fact that $\ket{s,m}$ are common eigenvectors of  $\hat{S}_{\rm a}$ and 
$\omega_{\rm a} \hat{S}_{\rm a}/2 + \hat{H}^{\rm aa}_{\rm int}$, with $\hat{H}^{\rm aa}_{\rm int}$ given in Eq.~\eqref{eq:Hint-Ising}. 

Since we have 
\begin{equation} 
\label{eq:epsa-pairs}
\hat\sigma_i^z \hat\sigma_{i+1}^z \ket{\alpha_i}\ket{\alpha_{i+1}} = 
\begin{cases}\begin{array}{cc}
+\ket{\alpha_i}\ket{\alpha_{i+1}}& \textrm{for}\ \alpha_i=\alpha_{i+1}, \\
-\ket{\alpha_i}\ket{\alpha_{i+1}}& \textrm{for}\ \alpha_i\neq\alpha_{i+1},
\end{array}\end{cases} 
\end{equation}  each summand in interaction Hamiltonian~\eqref{eq:Hint-Ising} multiplies $\ket{s,m}$ by either $1$ or $-1$. Thus, it holds that
\begin{equation} 
\label{eq:epsa-def}
\sum_{i=1}^N\hat\sigma_i^z \hat\sigma_{i+1}^z \ket{s,m}= \varepsilon_{\rm a}(s,m)\ket{s,m},
\end{equation} 
with $\varepsilon_{\rm a}(s,m)$ being an integer, which determines the interaction energy of source TLSs in the state $\ket{s,m}$.
From Eq.~\eqref{eq:epsa-pairs} it follows that the maximum value of $\varepsilon_{\rm a}$ is $N$. It is realized in a uniform state formed by a single homogeneous domain of source TLSs, all in the $\ket e$ or in the $\ket g$ state. Except for the case $\varepsilon_{\rm a}$ = $N$, the total number of homogeneous domains is even, with the number of $\ket{e}$ domains equal to the number of $\ket{g}$ domains. For example, a state with two pairs of $\alpha_i\neq \alpha_{i+1}$ (two domain walls) consists of two domains: one of $\ket e$ states, the other of $\ket g$ states, and has $\varepsilon_{\rm a}=N-4$. 

Let us assume that there exists at least one homogeneous $\ket{g}$-domain in the ring of source TLSs and denote by $d$ the number of homogeneous $\ket{e}$-domains with  
\begin{equation} 
d\in\{0,1,\dots,\lfloor N/2 \rfloor \}.
\end{equation} 
Here, $\lfloor N/2 \rfloor$ is the largest integer less than or equal to $N/2$. It gives the largest number of $\ket{e}$-domains if each domain is of length $1$. For a fixed $d$, there are $2d$ domain walls with $\alpha_i\neq \alpha_{i+1}$. The presence of each domain wall decreases $\varepsilon_{\rm a}$ by 2. Therefore, possible values of $\varepsilon_{\rm a}$ are 
\begin{equation} 
\label{eq:epsa-d}
\varepsilon_{\rm a}(d) = N-4d.
\end{equation} 
In this parametrization, $\varepsilon_{\rm a}(0)=N$ corresponds to the case of a single $\ket{g}$-domain. The same $\varepsilon_{\rm a}=N$ is obtained when there are no $\ket{g}$-domains at all. Hence, the case $\varepsilon_{\rm a}=N$
has the degeneracy two and corresponds to $2s=\pm N,d=0$.

Having introduced $s$, $\ket{s,m}$, and $\varepsilon_{\rm a}(d)$, we now find the weights $c_1$, $c_2$ in Eq.~\eqref{eq:alpha-definition} and the total energy $E$. By acting with $\hat{H}$ from Eq.~\eqref{eq:Htot-Ising-S} on $\ket\tau=\ket\alpha\ket{s,m}$, using Eqs.~\eqref{eq:Sa-eigenproblem} and~\eqref{eq:epsa-def}, and 
$\sigma_0^z \ket\alpha = c_1\ket{e}-c_2\ket{g}$, $\sigma_0^x \ket\alpha = c_2\ket{e}+c_1\ket{g}$, 
we get 
\begin{align} 
\hat{H} \ket{\alpha} \ket{s,m} & = \left[\frac{\omega_{0}}{2}( c_1\ket{e}-c_2\ket{g})+ \omega_{\rm a}s \ket{\alpha}+\gamma s(c_2\ket{e}+c_1\ket{g}) - \frac{J}{2}\varepsilon_{\rm a}(d)\ket{\alpha}\right] \ket{s,m}\nonumber\\
& = E \ket{\alpha} \ket{s,m}.
\end{align}  
This is equivalent to the eigenvalue problem in the matrix form 
\begin{equation}{\label{eq:alphainfull}} 
    \begin{pmatrix} 
        \frac{\omega_{0}}{2}+ \omega_{\rm a}s -\frac{J}{2}\varepsilon_{\rm a}(d) - E & \gamma s \\ 
        \gamma s & -\frac{\omega_0}{2}+\omega_{\rm a}s -\frac{J}{2}\varepsilon_{\rm a}(d) - E
    \end{pmatrix} 
    \begin{pmatrix} c_1 \\ c_2 \end{pmatrix}
    =\begin{pmatrix} 0 \\ 0  \end{pmatrix}, 
\end{equation} 
which yields a pair $E^\pm_{s,d}$ of the total system energies for given $s$ and $d$. They read  
\begin{equation} 
\label{eq:Epm} 
E^{\pm}_{s,d} = \omega_{\rm a}s - \frac{J}{2} (N-4d) \pm \frac{1}{2} \sqrt{\omega_0^2 + (2\gamma s)^2} \, .
\end{equation} 

Weights $c_{1,2}^\pm$ of the state~\eqref{eq:alpha-definition} corresponding to the energies $E^\pm_{s,d}$ are given by nontrivial solutions to the linear system  
\begin{subequations}
\label{eq:cs_equations}
\begin{align}
\label{eq:cs_1} 
&\left(\omega_0\mp\sqrt{\omega_0^2+(2\gamma s)^2} \right)c_1^\pm = -2 \gamma s\, c_2^\pm ,\\
\label{eq:cs_2} 
&-2\gamma s\, c_1^\pm =\left(-\omega_0\mp\sqrt{\omega_0^2+(2\gamma s)^2} \right) c_2^\pm.
\end{align}
\end{subequations} 
These equations determine $c_{1,2}^\pm$ as functions of the total magnetization $2s$ of source TLSs. In the matrix representation of the density operator $\hat\rho_0$, we will employ the following results: 
\begin{align} 
\label{eq:c12-solution}
& |c_2^\pm(s)|^2 = 1- |c_1^\pm(s)|^2= \frac{1}{2}\left(1\mp \frac{\omega_0}{\sqrt{\omega_0^2+(2\gamma s)^2}} \right), \\ 
\label{eq:c2c1}
& c_2^\pm(s)\left(c_1^\pm(s)\right)^* = \left(c_2^\pm(s)\right)^* c_1^\pm(s)=\pm \frac{\gamma s}{\sqrt{\omega_0^2+(2\gamma s)^2}}.
\end{align}

\section{Exact expression for the local quantum coherence} 

The partition function $Z(T,N)={\rm tr}\!\left[\exp(-\beta \hat H)\right]$ in Eq.~\eqref{eq:def_totalrho} is the sum over all eigenstates of $\hat H$,  
\begin{equation}
\label{eq:Z}
Z(T,N)=\displaystyle\sum_{s=-N/2}^{N/2}\displaystyle\sum_{d=0}^{\lfloor N/2 \rfloor}  \Omega(s, d, N)\left(e^{-\beta E^+_{s,d}} + e^{-\beta E^-_{s,d}}\right), 
\end{equation}
where energies $E^{\pm}_{s,d}$ are given in Eq.~\eqref{eq:Epm} and the degeneracy~\cite{Antal/etal:2004}
\begin{equation} 
\label{eq:Omega-result}
\Omega(s,d,N) = \begin{cases} \delta_{|2s|,N}\, \delta_{d,0} & \textrm{for}\ |2s|=N\;\text{or}\; d=0, \\[2ex] \displaystyle \frac{N}{d} \binom{N/2-|s|-1}{d-1}\binom{N/2+|s|-1}{d-1}&\textrm{for}\ |2s|\neq N\; \text{and} \;d\neq 0, \end{cases} 
\end{equation} 
equals the number of microstates of source TLSs for a fixed $s$, $d$, and $N$.
In Eq.~\eqref{eq:Omega-result}, $\delta_{i,j}$ denotes the Kronecker delta. 
The product of binomial coefficients vanishes if $d>N/2-|s|$.
For the derivation of Eq.~\eqref{eq:Omega-result} and further combinatorial results related to the one-dimensional Ising model with various boundary conditions, see Refs.~\cite{Antal/etal:2004, Denisov/Hanggi:2005, Plascak:2018, Dantchev/Rudnick:2022}. 

The equilibrium density operator $\hat\rho$ of the entire system is conveniently represented in the basis $\{\ket{\alpha^+(s)}\ket{s,m}$, $\ket{\alpha^-(s)}\ket{s,m} \}$, where the target TLS state $\ket{\alpha^\pm}$ is defined in Eq.~\eqref{eq:alpha-definition}. Tracing over the state space of the source TLSs gives us the density operator~\eqref{eq:rho0-def} of the target TLS: 
\begin{equation}
\hat\rho_0 = \frac{1}{Z} \sum_{s,d} \Omega(s,d,N) 
\left( e^{-\beta E^+_{s,d}} \ket{\alpha^+(s)}\bra{\alpha^+(s)} +e^{-\beta E^-_{s,d}} \ket{\alpha^-(s)}\bra{\alpha^-(s)}  \right).
\end{equation}
Elements of its matrix representation~\eqref{eq:rho0-def-matrix} over the bare basis $\{ \ket e,\ket g \}$ read 
\begin{subequations}
\label{eq:elementsrho0}
\begin{align} 
& \rho_{g} = 1-\rho_{e} =\frac{1}{Z} \sum_{s,d} \Omega(s,d,N) \left(|c_2^+(s)|^2 e^{-\beta E^+_{s,d}} + |c_2^-(s)|^2 e^{-\beta E^-_{s,d}} \right) ,\\
\label{eq:rhoge}
& \rho_{ge} = \rho_{eg}^* = \frac{1}{Z} \sum_{s,d} \Omega(s,d,N)\, \left(c^+_2(s)c_1^+(s)^* e^{-\beta E^+_{s,d}} +c^-_2(s)c_1^-(s)^* e^{-\beta E^-_{s,d}}  \right).
\end{align} 
\end{subequations}
Using Eqs.~\eqref{eq:Epm} and~\eqref{eq:c2c1}, we get the exact expression for the coherence~\eqref{eq:C-definition},  
\begin{equation}
\label{eq:C-exact}
C = \left| \frac{4 \gamma}{Z}\sum_{s,d} \frac{s\,\Omega(s,d,N) }{\sqrt{\omega_0^2+(2\gamma s)^2}} e^{-\beta\left[ \omega_{\rm a}s - J(N-4d)/2\right]} \sinh(\frac{\beta}{2} \sqrt{\omega_0^2 + (2\gamma s)^2}) \right|, 
\end{equation} 
which can be further simplified by rewriting it as a sum over $s>0$. For $\omega_{\rm a}>0$, we have 
\begin{equation} 
\label{eq:C-R} 
C = \frac{8 |\gamma|}{Z}\sum_{s>0} R(s,N) \frac{s \sinh(\beta \omega_{\rm a} s)}{\sqrt{\omega_0^2+(2\gamma s)^2}}  \sinh(\frac{\beta}{2} \sqrt{\omega_0^2 + (2\gamma s)^2}), 
\end{equation}
where the weighting factor is 
\begin{align} 
\label{eq:R} 
R(s,N) = \displaystyle\sum_{d=0}^{N/2-s} \Omega(s,d,N)\, e^{\beta J(N-4d)/2}.
\end{align} 
It is proportional to the partition function of $N$ source TLSs conditioned on a sharp value of the total magnetization $2s$. 
We rewrite the partition function~\eqref{eq:Z} analogously: 
\begin{align} 
\label{eq:Z-R} 
\begin{split} 
Z = 4 \sum_{s> 0} R(s,N) \cosh(\beta \omega_{\rm a} s)\cosh(\frac{\beta}{2} \sqrt{\omega_0^2 + (2\gamma s)^2})  + \begin{cases} 2 R(0,N) \cosh(\frac{\beta \omega_0}{2}) & \textrm{for}\ N\ \text{even},\\ \displaystyle 0  &\textrm{for}\ N\ \text{odd}. \end{cases}
\end{split}
\end{align} 
Notably, $C$ and $Z$ are given by Eqs.~\eqref{eq:C-R} and~\eqref{eq:Z-R}, but with different $R(s,N)$, for a certain class of interactions between source TLSs, see App.~\ref{sec:appendixA}. 

In Secs.~\ref{sec:C-general}-\ref{sec:C-antiferromagnetic-odd}, we discuss the temperature dependence of $C$ and how the shape of the function $C(T)$ is affected by values of the model parameters $J$, $\omega_{\rm a}$, $\gamma$, and $N$. Since $C$ depends on $|\gamma|$, we focus on $\gamma>0$ and also assume that $\omega_{\rm a}>0$ and $\omega_0 > 0$. In the illustrations, we choose units such that $k_{\rm B}=1$.

\section{Coherence growing with \texorpdfstring{$J$}{\textit{J}}, \texorpdfstring{$\omega_{\rm a}$}{omega{\scriptsize a}}, and \texorpdfstring{$\gamma$}{gamma}, upper and lower bounds on \texorpdfstring{$C(T)$}{\textit{C(T)}}} 
\label{sec:C-general} 

Variation of model parameters can lead to a qualitative change of the function $C(T)$, transforming a monotonically decreasing function (for $J>0$, see Sec.~\ref{sec:C-ferromagnetic}) into one attaining a maximum at a certain finite temperature due to a ground-state level crossing (Secs.~\ref{sec:C-antiferromagnetic-even} and~\ref{sec:C-antiferromagnetic-odd}). Before exploring this rich behavior in detail, let us interpret the inequalities 
\begin{align}
\label{eq:C-inequality-J}
& C(T,J_1) < C(T,J_2)
\quad \textrm{for} \quad  -\infty<J_1 < J_2<+\infty,\\
\label{eq:C-inequality-wa}
& C(T,\omega_{\rm a1}) < C(T,\omega_{\rm a2})
\quad \textrm{for} \quad  0<\omega_{\rm a1} < \omega_{\rm a2}<+\infty, \\ 
\label{eq:C-inequality-gamma}
& C(T,\gamma_1) < C(T,\gamma_{2})
\quad \textrm{for} \quad  0<\gamma_{1} < \gamma_{2} <\infty,
\end{align}
which hold for $T>0$ and are proven in Apps.~\ref{sec:appendix-C-monotonic}-\ref{sec:appendix-C-monotonic_gamma}. The notation $C(T, J)$, $C(T, \omega_{\rm a})$, and $C(T,\gamma)$ displays the dependence of coherence~$C$ on~$T$ and~$J$,~$\omega_{\rm a}$, or~$\gamma$. 

According to inequality~\eqref{eq:C-inequality-J}, at arbitrary $T>0$, the coherence is a monotone increasing function of~$J$. Thus, the source TLSs with ferromagnetic interaction ($J>0$) can generate a larger $C$ than noninteracting ones ($J=0$) and than those with antiferromagnetic interaction ($J<0$). 
Consequently, the upper bound $C_{\rm ub}(T)$ on the coherence at a given temperature, 
\begin{equation}
C(T,J) < C_{\rm ub}(T), 
\end{equation}
is obtained by taking the $J\to\infty$ limit in Eq.~\eqref{eq:C-R}. 
To this end, note that for $NJ/k_{\rm B}T \gg 1$, the sums in Eq.~\eqref{eq:C-R} are dominated by the exponential terms with the largest multiple of $J$, which occurs for $d=0$ and $s=N/2$. That is, as $J\to\infty$, $E^\pm_{s,d}\to-J(N-4d)/2$ and the gaps between levels with different $d$ diverge. Therefore, the system can occupy only states with minimal $d$, which are $s=\pm N/2$, $d=0$ states, corresponding to the last terms in the sum~\eqref{eq:C-R}.   
Hence, we get
\begin{equation}
\label{eq:C-upperbound}
C_{\rm ub}(T)=\frac{\gamma N}{\sqrt{\omega_0^2+(\gamma N)^2}}\tanh\left(\frac{\beta N\omega_{\rm a}}{2}\right)\tanh\left(\frac{\beta}{2}\sqrt{\omega_0^2+(\gamma N)^2}\right).
\end{equation} 
For $N=1$, $C_{\rm ub}(T)$ is identical to $C$ generated by a single source TLS. The strong ferromagnetic interaction thus effectively joins the source TLSs into one with the energy gap $N\omega_{\rm a}$. This joint system interacts with the target TLS via the effective coupling constant $N\gamma$.

If the interaction between the source TLSs is antiferromagnetic and strong compared to the thermal energy ($NJ/k_{\rm B}T\ll -1$), $C$ becomes greatly reduced. The lower bound $C_{\rm lb}(T)$,
\begin{equation} 
C_{\rm lb}(T) <  C(T,J),
\end{equation}
obtained as the $J\to -\infty$ limit taken in Eq.~\eqref{eq:C-R}, can be different from zero. Whether it is zero or not depends on the parity of $N$. 

If $N$ is odd and $J\to-\infty$, the sums in Eq.~\eqref{eq:C-R} are dominated by terms with the greatest $d$, which are $s=1/2,d=(N-1)/2$ terms in both numerator and denominator. The lower bound on $C$ is then nonzero: 
\begin{equation}
\label{eq:Clb-odd} 
C_{\rm lb}(T) = 
\frac{\gamma }{\sqrt{\omega_{0}^2+\gamma^2}} \tanh\!\left(\frac{\beta \omega_{\rm a}}{2}\right) \tanh{\!\left(\frac{\beta}{2}\sqrt{\omega_{0}^2+\gamma^2 }\right)}, 
\quad N\ \textrm{odd}.
\end{equation}
Similarly to $C_{\rm ub}(T)$ in Eq.~\eqref{eq:C-upperbound}, this result is identical to the coherence induced by a single source TLS. The strong antiferromagnetic interaction diminishes $s$ (and hence $C$) by arranging $\lfloor N/2\rfloor$ source TLSs into $\ket e$ and $\lfloor N/2\rfloor$ into $\ket g$ states with a maximum number of domain walls. The paired TLSs effectively compensate their effects on the total magnetization $2s$. The coherence $C_{\rm lb}(T)$ is then induced by $s=1/2$ corresponding to a single unpaired source TLS. 

For even $N$ and $J \to -\infty$, the sums in Eq.~\eqref{eq:C-R} are again dominated by terms corresponding to the highest values of $d$. In the numerator, the dominant contribution arises from the term with $s = 1$ and $d = N/2 - 1$, while the partition function $Z$ in the denominator includes an additional term with $s=0$ and $d=N/2$, in contrast to the case of odd $N$, cf. Eq.~\eqref{eq:Z-R}. As a result, $C \sim e^{2\beta J}$, yielding
\begin{equation}
\label{eq:Clb-even} 
C_{\rm lb}(T) = 0, 
\quad N\ \textrm{even}.
\end{equation}
Here, the strong antiferromagnetic interaction antialigns all source TLSs, implying $s=0$, and there is no induced coherence in the target TLS.

\section{Temperature dependence of \texorpdfstring{$C$}{{\it C}} for ferromagnetic interaction \texorpdfstring{$J>0$}{ }} 
\label{sec:C-ferromagnetic}  

\begin{figure}[t!]
\center{\includegraphics[width=1\linewidth]{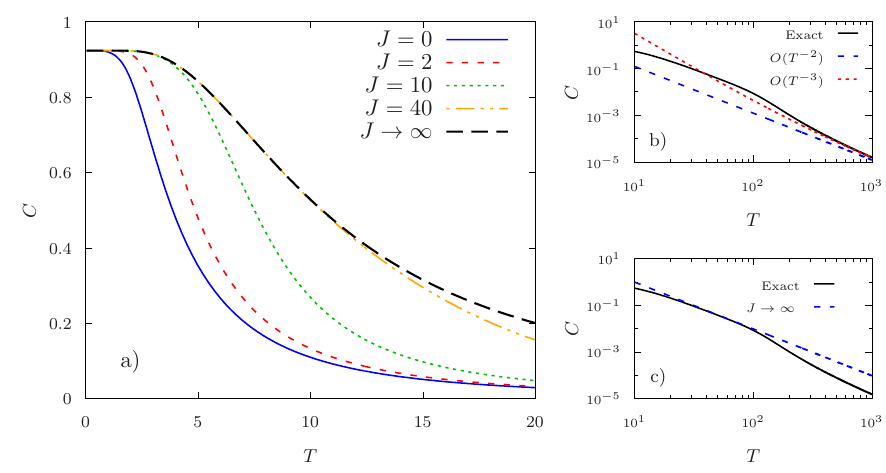}}
\caption{a)~$C$ [Eq.~\eqref{eq:C-R}] as a function of temperature $T$ for different values of $J\geq 0$. Curves for finite $J>0$ lie between the upper bound~\eqref{eq:C-upperbound} ($J\to\infty$, dashed black line) and $C$ for $J=0$ (solid blue line). 
b)~The lowest-order $O(T^{-2})$ term in the asymptotic approximation~\eqref{eq:C-highT} deviates from the exact result~\eqref{eq:C-R}. 
c)~Limits $T\to\infty$ and $J\to\infty$ do not commute. Parameters used: $\omega_{0}=10$, $\omega_{\rm a}=2$, $\gamma=3$, $N=8$, and in b) and c) $J=250$.
The figure shows that the ferromagnetic coupling between the sources ($J>0$) leaves the zero-temperature coherence unchanged and suppresses its thermal decay. As $J$ increases, the low-temperature plateau extends toward the strong-coupling limit, in which the sources behave as a single collective spin. The coupling also enters a subleading $\sim T^{-3}$ term to the high-temperature decay, leading to noncommuting limits as $T\to\infty$ and $J\to\infty$.}
\label{fig:fig1_posJonJ} 
\end{figure}

Figure~\ref{fig:fig1_posJonJ}~a) shows the coherence $C$ as a function of $T$ for various values of $J\geq 0$. We have chosen the bare gap of target TLS $\omega_0=10$, hence $T=10$ corresponds to the thermal energy comparable to that gap. The upper bound~\eqref{eq:C-upperbound} is represented by the black dashed line. The lower bound for $J\geq 0$ (solid blue line) is $C(T)$ evaluated at $J=0$, i.e., noninteracting source TLSs. 
 
In agreement with the inequality~\eqref{eq:C-inequality-J}, the increase in $J$ has an overall positive effect on the coherence. In this respect, Fig.~\ref{fig:fig1_posJonJ} reveals interesting aspects: 
$C$ is highest in the $T\to 0$ limit. This maximal coherence appears to be independent of $J$. Increasing $J$ extends the plateau of nearly-constant $C$ values at low temperatures.  The value of $J$ also affects the high-temperature algebraic decay of $C$. 

In fact, the underlying behavior of $C(T,J)$ is even richer. We shall address the individual temperature regimes by means of analytical calculations in Secs.~\ref{sec:C-T0-ferro} and~\ref{sec:C-highT-ferro}. Moreover, the dependence of $C$ on $\gamma$, $\omega_{\rm a}$, and $N$ will be detailed in Secs.~\ref{sec:C-gamma-ferro}-\ref{sec:C-N-ferro}, respectively, and illustrated in Fig.~\ref{fig:fig2_posJ_onpar}.

\subsection{Coherence at \texorpdfstring{$T=0$}{zero temperature}}
\label{sec:C-T0-ferro}
Let us find the ground state of the entire system for $J>0$. Equation~\eqref{eq:Epm} implies that 
$E^{-}_{s,d}<E^{+}_{s,d}$ holds and that for $J>0$ and fixed $s$, $E^-_{s,d}$ is an increasing function of $d$. 
Depending on the total magnetization $2s$, there are two values of $d$ for which $E^-_{s,d}$ is minimized: 
(i) if $2s=\pm N$, we obtain the minimum energy for $2d=0$, i.e., no domain walls occur in the ground state, corresponding to a homogeneous domain of source TLSs;
(ii) if $N>2s>-N$, we get the minimum energy $2d=2$, corresponding to two homogeneous domains. 
Moreover, in case of $\omega_{\rm a}>0$, Eq.~\eqref{eq:Epm} yields 
$E^{-}_{s,1}>E^{-}_{-N/2,0}$, for $-N< 2s< N$. That is, due to $\omega_{\rm a}>0$, microstates with $2s>-N$ have a higher lowest energy compared to the one with $s=-N/2$. 
The ground state energy of the entire system thus reads  
\begin{equation} 
\min_{s,d}\{E^+_{s,d},E^-_{s,d}\} = E^-_{-N/2, 0}\, ,\quad \textrm{for}\ J>0.
\end{equation}

This means that at $T=0$ the entire system resides in the ground state with $s=-N/2$, $d=0$, $\Omega(-N/2,0,N)=1$, and the ground state coherence is  
\begin{equation} 
\label{eq:C0-result}
C_0 = \lim_{T\to 0} C(T) = \frac{\gamma  N}{\sqrt{\omega_0^2+(\gamma N)^2}},\quad J>0.
\end{equation} 
$C_0$ decreases with increasing $\omega_0$ and it grows with $\gamma N$, monotonically approaching the upper bound 1 as $\gamma N\to \infty$.  
Notably, the ground state coherence does not depend on the gap $\omega_{\rm a}$ of source TLSs. It is independent of $J$ also, yet in its derivation we assumed that $J>0$.

With $\omega_{0}=10$ and $\gamma N=24$ in Fig.~\ref{fig:fig1_posJonJ}, the ground-state coherence $C_0\simeq 0.923$ is rather high. Other values of $C_0$ can be seen in Fig.~\ref{fig:fig2_posJ_onpar} below.

\subsection{High-temperature decay of \texorpdfstring{$C(T)$}{coherence}}
\label{sec:C-highT-ferro}
As illustrated in Fig.~\ref{fig:fig1_posJonJ}, the coherence vanishes with increasing temperature. When $T\to\infty$, all $2^{N+1}$ eigenstates of the total Hamiltonian become equally probable and $e^{-\beta E^\pm_{s,d}}\to 1$ in Eqs.~\eqref{eq:elementsrho0}. Then, $\rho_{ge}$ and $\rho_{eg}$ in Eq.~\eqref{eq:rhoge} approach $0$, since terms with $s$ and $-s$  cancel each other due to   $c_2^\pm(s)\left(c_1^\pm(s)\right)^* =-c_2^\pm(-s)\left(c_1^\pm(-s)\right)^*$, cf.\ Eq.~\eqref{eq:c2c1}.

How fast does the coherence decay toward zero with increasing temperature? The result of the expansion of Eq.~\eqref{eq:C-R} into a power series in $\beta$ turns out to be  
\begin{equation}
\label{eq:C-highT}
C(T)\sim \frac{\omega_{\rm a} \gamma N}{4T^2}\left(1+\frac{J}{T}\right), \quad T\to\infty .
\end{equation} 
The leading $O(T^{-2})$ term of the asymptotic approximation does not depend on $J$ and represents the overall effect of the interaction between the target TLS and $N$ independent source TLSs. The next $O(T^{-3})$ term results from an average interaction of the target TLS with independent pairs of interacting source TLSs. 

Increasing $J$ enhances the coherence. In this regard, the asymptotics~\eqref{eq:C-highT} behave in accordance with the inequality~\eqref{eq:C-inequality-J}.  Since the decay with $T$ is algebraic, the interaction-induced $O(T^{-3})$ term can be significant even at high temperatures. This is demonstrated in a log-log graph in Fig.~\ref{fig:fig1_posJonJ}~b) showing the exact curve $C(T)$, the main $O(T^{-2})$ asymptotics only, and both terms of approximation~\eqref{eq:C-highT}.

One may also wonder whether an asymptotic approximation of $C(T)$ can be derived from the simple expression~\eqref{eq:C-upperbound} for the upper bound. Such an approach yields  
\begin{equation} 
C_{\rm ub}(T) \sim \frac{\omega_{\rm a}\gamma N^2}{4T^2},\quad T\to\infty ,
\end{equation} 
with a different power of $N$ than in~\eqref{eq:C-highT}. 
While in~\eqref{eq:C-highT} the target TLS interacts with $N$ independent source TLSs of gap $\omega_{\rm a}$ and the interaction strength is $\gamma$, here, it interacts with a single effective TLS of gap $N\omega_{\rm a}$ by the interaction of strength $\gamma N$, cf.\ the interpretation of Eq.~\eqref{eq:C-upperbound}. Thus, the limits $T\to\infty$ and $J\to\infty$ do not commute, which is illustrated in Fig.~\ref{fig:fig1_posJonJ}~c).

\subsection{Effect of \texorpdfstring{$\gamma$}{gamma}}
\label{sec:C-gamma-ferro} 

\begin{figure}[t!]
\center{\includegraphics[width=1\linewidth]{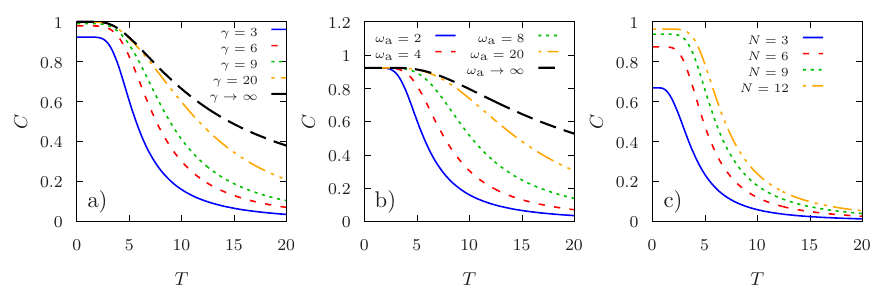}}
\caption{a)~Increasing $\gamma$ slows down the decay of $C$ with $T$ and increases $C$ near $T=0$. The black dashed line represents the $\gamma\to\infty$ limit~\eqref{eq:C-highgamma}. 
b)~Increasing $\omega_{\rm a}$ yields higher coherence at finite $T$ up to the upper bound given in Eq.~\eqref{eq:C-highwa} (black dashed line). $C(0)$ does not change with $\omega_{\rm a}$, see Eq.~\eqref{eq:C0-result}.  
c)~$C$ at any $T$ is an increasing function of $N$. Parameters used: $\omega_{0}=10$, $J=4$, and a) $\omega_{\rm a}=2$, $N=8$, b) $\gamma=3$, $N=7$, c) $\omega_{\rm a}=2$, $\gamma=3$. The figure illustrates two distinct control mechanisms: Increasing the coupling $\gamma$ or the number of sources $N$ enhances coherence at all temperatures, including $T=0$. By contrast, increasing the source gap $\omega_a$, like increasing $J$, improves the resilience of $C(T)$ to temperature without changing $C(0)$.}
\label{fig:fig2_posJ_onpar}
\end{figure}

Figure~\ref{fig:fig2_posJ_onpar}~a) demonstrates how changing the interaction strength $\gamma$ affects the function $C(T)$. In agreement with Eq.~\eqref{eq:C0-result}, larger $\gamma$ yields higher coherence in the $T\to 0$ limit. A similar positive effect occurs also at nonzero $T$, cf.\ the inequality~\eqref{eq:C-inequality-gamma}. The upper bound for the $\gamma$-induced enhancement of $C(T)$ reads 
\begin{equation}
\label{eq:C-highgamma}
\lim_{\gamma\to\infty} C(T,\gamma) = \tanh{\!\left(\frac{\beta N\omega_{\rm a}}{2}\right)}. 
\end{equation} 
This ultra-strong coupling limit is represented by the black dashed line in Fig.~\ref{fig:fig2_posJ_onpar}~a). 

In the weak-coupling regime, the coherence vanishes linearly with~$\gamma$:
\begin{equation}
\label{eq:smallgamma} 
C(T,\gamma)\sim\frac{2\gamma}{\omega_0}\tanh{\!\left(\frac{\beta\omega_0}{2}\right)}\frac{\displaystyle\sum_{s>0}sR(s,N)\sinh(\beta s \omega_{\rm a})}{Q(\beta,\omega_{\rm a},J,N)},\quad \gamma\to 0,
\end{equation} where 
\begin{equation}
Q(\beta,\omega_{\rm a},J,N)=\sum_{s>0}R(s,N)\cosh(\beta s \omega_{\rm a}) 
+\begin{cases} R(0,N)/2\quad&\text{for $N$ even,} \\
0\quad&\text{for $N$ odd.}
\end{cases}
\end{equation}

\subsection{Effect of \texorpdfstring{$\omega_{\rm a}$}{omega{\scriptsize a}}}
As illustrated in Fig.~\ref{fig:fig2_posJ_onpar}~b), another parameter that can increase the coherence at nonzero $T$ is the gap size $\omega_{\rm a}$. The upper bound, see Eq.~\eqref{eq:C-inequality-wa}, limiting this increase is given by 
\begin{align}
\label{eq:C-highwa}
\lim_{\omega_{\rm a}\to \infty} C(T,\omega_{\rm a})=\frac{\gamma N}{\sqrt{\omega_0^2+(\gamma N)^2}}\tanh{\left(\frac{\beta}{2}\sqrt{\omega_0^2+(\gamma N)^2}\right)}.
\end{align}
It corresponds to the terms with  $s=N/2$ in the sums of Eq.~\eqref{eq:C-R} as they give major contributions to the sums in the $\omega_{\rm a}\to\infty$ limit.

In the small-gap limit ($\omega_{\rm a}\to 0$), the coherence vanishes, i.e., degenerate source TLSs do not induce the local coherence. 

Let us compare the effects of $\omega_{\rm a}$ [Fig.~\ref{fig:fig2_posJ_onpar}~b)] and $J\geq 0$ [Fig.~\ref{fig:fig1_posJonJ}~a)] on the autonomous coherence and relate them to the microstates and the spectrum of the system. At $T=0$, where $C$ attains its global maximum~\eqref{eq:C0-result}, there is no impact of $\omega_{\rm a}$ or $J$ on the value of $C$. An increase in $\omega_{\rm a}$ or $J$ leads to the enhancement of $C$ at finite temperatures. Increasing these parameters also extends the temperature interval of small $T$ values over which $C$ remains approximately constant, thus postponing the temperature-induced decay of the coherence. In the high-temperature regime, see Eq.~\eqref{eq:C-highT}, the leading-order contribution to $C$ scales with $\omega_{\rm a}$, whereas $J$ appears only in the subleading correction. This asymmetry points to distinct roles played by the two parameters when determining the statistical weights of the system's microstates. 

Both $\omega_{\rm a}$ and $J$ enter the energy spectrum~\eqref{eq:Epm} additively. The term proportional to $\omega_{\rm a}$ couples with the total magnetization $2s$ of the environment. The contribution from $J$ depends on the number $d$ of domain walls. The gap $\omega_{\rm a}$ determines the energy scale in the environment. If $\omega_{\rm a}$ is large compared to the thermal energy, the magnetization of the environment is high as all the source TLSs most likely reside in the ground state for any $J\geq 0$. On the other hand, the impact of $J$ is related to the local correlations of the source TLSs and determines the degree of partial local ordering within the environment. As a result, the increase in both parameters enhances the magnetization of the environment, and therefore in certain aspects they have an analogous impact on $C$, which is generated via the magnetization-dependent interaction, see Eq.~\eqref{eq:Htot-Ising-S}. 

To conclude, increasing the parameters $\omega_{\rm a}$ and/or $J$ can provide a practical method to enhance the coherence at finite $T$, which may facilitate the experimental detection of $C$. Such generated robustness against the thermal noise is particularly relevant in scenarios where the temperature control is limited, yet the coherence must be preserved.

\subsection{Effect of \texorpdfstring{$N$}{{\it N}}} 
\label{sec:C-N-ferro}
Figure~\ref{fig:fig2_posJ_onpar}~c) demonstrates the enhancement of $C(T)$ with increasing $N$. The ground-state ($T=0$) coherence grows with $N$ according to Eq.~\eqref{eq:C0-result}. The larger $N$ also yields a more extended plateau observed near $T=0$. This must be due to the widening of the energy gap between the ground state and the first excited state. The large-temperature tail~\eqref{eq:C-highT} increases with $N$ as well. In the entire temperature range, $C\to 1$ as $N\to\infty$, which is confirmed by the derived large-$N$ asymptotic behavior of $C$ in App.~\ref{sec:appendix-D-large-N}. Thus, $N$ is an important parameter for enhancing the coherence globally.

\section{Temperature dependence of \texorpdfstring{$C$}{{\it C}} for antiferromagnetic interaction \texorpdfstring{($J<0$)}{ } and even \texorpdfstring{$N$}{{\it N}}}
\label{sec:C-antiferromagnetic-even}

For antiferromagnetic interaction ($J<0$) and even number $N$ of source TLSs, one may expect the ground state to have the maximum number $2d=N$ of domain walls and zero total magnetization $s=0$, which corresponds to a perfectly antialigned configuration with $N/2$ TLSs in state $\ket g$ and $N/2$ in $\ket e$. This expectation turns out to be true for a limited range of model parameters. The derivation in Sec.~\ref{sec:QPT-Neven} below demonstrates the second possibility, where the ground state is doubly degenerate with $d=0$ (no domain walls) and $s=-N/2$, corresponding to fully aligned source TLSs.

\begin{figure}[ht!]
\centering 
\includegraphics[width=1\linewidth]{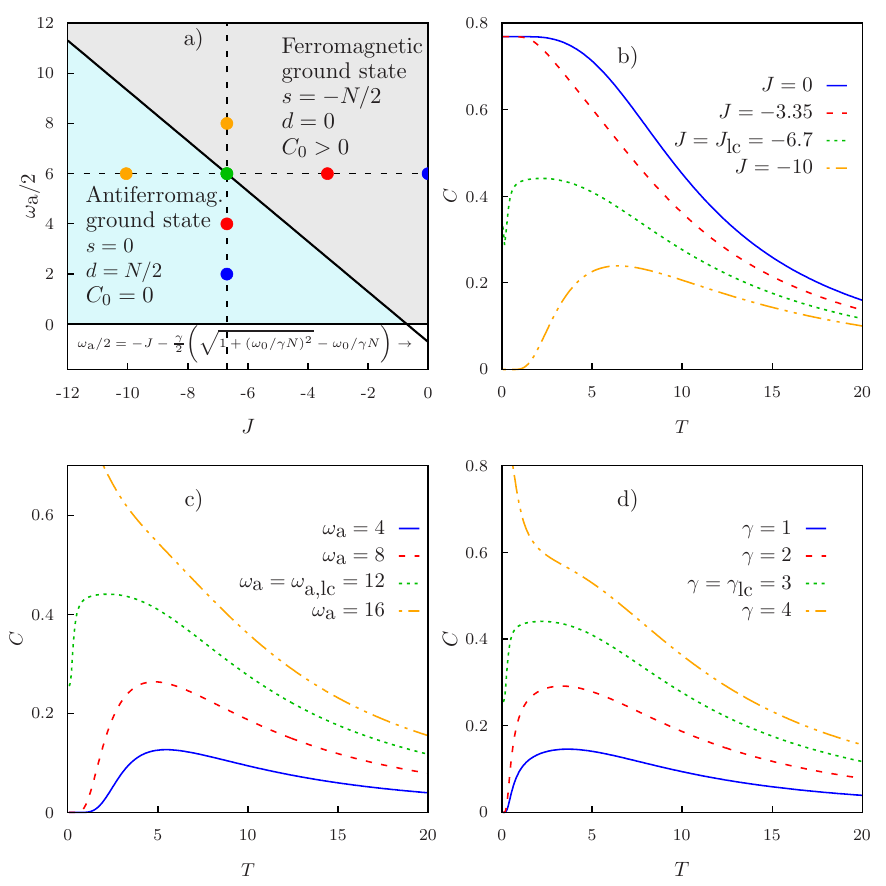}
\caption{a)~The phase diagram for fixed $\omega_0=20$, $\gamma=3$, and even $N=8$. In the gray (upper) region, the ground state is ferromagnetic with $C_0>0$, while in the blue region it is antiferromagnetic and $C_0=0$. The ground-state level crossing line follows the equation shown within the white region (bottom). Coordinates $(J,\omega_{\rm a}/2)$ of the colored circles on a horizontal dashed line correspond to the parameter values used to draw the curves of the respective color in b). The circles on the vertical line correspond to the curves in c). b)~$C(T)$ for different values of the interaction strength $J$, illustrating typical behavior of $C(T)$ for the two phases and on the ground-state level crossing line. c)~$C(T)$ for different values of the source TLSs gap $\omega_{\rm a}$. d)~$C(T)$ for different values of the coupling constant $\gamma$ and for $\omega_{\rm a}=12$, $J=-6.7$, $N=8$. Increasing $\omega_{\rm a}$ has a qualitatively similar effect to increasing $\gamma$: it slows the decrease of $C(T)$ and raises its maximum value. The figure illustrates that for antiferromagnetic coupling $J<0$, the target-source interaction can drive a ground-state crossover from an aligned to a N\'{e}el-like antiferromagnetic source configuration. In the N\'{e}el-like regime, the coherence vanishes at $T=0$ but develops a maximum at finite temperature, as thermal population of magnetized states induces a tilt of the target. Thus, in this regime, temperature seems to generate coherence rather than merely suppressing it.}
\label{fig:fig3_negJ_evenN} 
\end{figure}

The transition between these two types of ground state is represented in $(\omega_{\rm a}/2)J$-plane in Fig.~\ref{fig:fig3_negJ_evenN}~a) with regions of $s=0$ and $s=-N/2$ separated by the linear function  
\begin{equation}
\label{eq:Jc-Neven}
J_{\rm lc}+\frac{\omega_{\rm a,lc}}{2}+\frac{\gamma_{\rm lc}}{2}\left( \sqrt{1+\left(\frac{\omega_{0,{\rm lc}}}{\gamma_{\rm lc} N_{\rm lc}}\right)^2} -\frac{\omega_{0,{\rm lc}}}{\gamma_{\rm lc} N_{\rm lc}} \right) = 0, 
\end{equation} where $J_{\rm lc},\gamma_{\rm lc},\omega_{0,{\rm lc}},\omega_{\rm a,lc},$ and $N_{\rm lc}$ are the values of model parameters for which the ground-state level crossing occurs.

The two types of ground state yield different coherences at $T=0$:  
\begin{equation} 
\label{eq:C0-Jnegative-Neven} 
C_0 = \begin{cases}  
0  & \textrm{for}\ s=0,\ d=N/2, \\[0.5ex]
\displaystyle \frac{\gamma N}{\sqrt{\omega_0^2+ (\gamma N)^2}} & \textrm{for}\ s=-N/2,\ d=0. 
\end{cases}
\end{equation} 
On the ground-state level crossing line~\eqref{eq:Jc-Neven}, the ground state is degenerate as the energies of $s=0$ and $s=-N/2$ states become equal. The zero-temperature coherence then reads 
\begin{equation} 
\label{eq:C0-Jnegative-Neven-pt} 
C_0 =\frac{1}{3} \frac{\gamma N}{\sqrt{\omega_0^2+ (\gamma N)^2}}. 
\end{equation} 
The shape of $C(T)$ is strongly influenced by $C_0$. As a result, the presence of the ground-state level crossing leads to pronounced changes in $C(T)$ even at nonzero temperatures, see Figs.~\ref{fig:fig3_negJ_evenN}~b)-d). 

Figure~\ref{fig:fig3_negJ_evenN}~b) shows the coherence $C(T)$ for various values of $J$ located on the horizontal dashed line of Fig.~\ref{fig:fig3_negJ_evenN}~a), indicated by circles with the corresponding colors. The transition point $J_{\rm lc}$ is calculated by solving Eq.~\eqref{eq:Jc-Neven} for $\omega_0=20$, $\gamma=3$, $\omega_{\rm a}=12$, and $N=8$. For $J\in(J_{\rm lc},0)$, the ground state is characterized by $s=-N/2$, $d=0$, and $C(T)$ exhibits a global $J$-independent maximum at $T=0$, similarly to the case of ferromagnetic interaction with $J>0$ shown in Fig.~\ref{fig:fig1_posJonJ}. For $J<J_{\rm lc}$, the ground state has $s=0$ and $d=N/2$, the curve $C(T)$ starts from $C_0=0$ and reaches a maximum at some nonzero~$T$. At the ground-state level crossing line~\eqref{eq:Jc-Neven}, the ground state coherence is nonzero, see Eq.~\eqref{eq:C0-Jnegative-Neven-pt}, and a maximum of $C(T)$ can occur at some $T>0$. The existence of this maximum stems from the large coherence of the first excited state relative to the ground state. 

An analogous correspondence between the type of the ground state and the shape of $C(T)$ is observed in Fig.~\ref{fig:fig3_negJ_evenN}~c), where $C(T)$ is plotted for various $\omega_{\rm a}$. Varying $\omega_{\rm a}$ in this case corresponds to passing the ground-state level crossing point on the vertical dashed line in Fig.~\ref{fig:fig3_negJ_evenN}~a).

Finally, Fig.~\ref{fig:fig3_negJ_evenN}~d) shows $C(T)$ for different interaction strengths $\gamma$. The increase in $\gamma$ corresponds to the vertical downward shift of the ground-state level crossing line in Fig.~\ref{fig:fig3_negJ_evenN}~a) towards smaller values of $\omega_{\rm a}$. Increasing $\gamma$ enhances $C(T)$ at all $T>0$. In the strong-coupling limit $\gamma\to \infty$, $C(T)$ converges to the limiting form~\eqref{eq:C-highgamma}. In the weak-coupling limit $\gamma\to 0$, we get $C(T)=0$.

\subsection{Ground state energy level crossing} 
\label{sec:QPT-Neven}

Equation~\eqref{eq:Epm} implies that $E^{-}_{s,d}<E^{+}_{s,d}$ holds, therefore, we find the ground state energy by minimizing $E^-_{s,d}$ with respect to $s$ and $d$. For $J<0$, the energy $E^{-}_{s,d}$ is a decreasing function of $d$. The smallest energy for a given $s$ is reached for the largest possible $d$, which is $d=N/2-|s|$, see Eq.~\eqref{eq:Omega-result}. Minimum value of energy $E^{\rm min}_{s}=E^-_{s,N/2-|s|}$ at a fixed $s$ satisfies $E^{\rm min}_{-|s|}< E^{\rm min}_{+|s|}$ if $s\neq 0$ and $\omega_{\rm a}>0$, hence the minimum energy as a function of $s$ reads  
\begin{equation}
\label{eq:minimumE} 
E^{\rm min}_{s} = \omega_{\rm a} s+ J\left(\frac{N}{2}+2s\right) - \frac{1}{2} \sqrt{\omega_{0}^2+(2\gamma s)^2} ,
\end{equation} 
where $s\in\{-N/2,\ldots,0\}$. Analyzing the derivative $\partial E^{\rm min}_{s}/\partial s$, $s\in [-N/2,0]$ and the values $E^{\rm min}_{-N/2}$ and $E^{\rm min}_{0}$ leads to the conclusion that the minimum energy occurs at the boundary value $s=0$ or $s=-N/2$. Which of the two energy levels corresponds to the global minimum depends on the values of $\omega_0$, $\omega_{\rm a}$, $J$, $\gamma$, and $N$. If we introduce the energy difference 
\begin{equation} 
\Delta E^{\rm min} = E^{\rm min}_{0}-E^{\rm min}_{-N/2}, 
\end{equation} 
the level crossing occurs at $\Delta E^{\rm min}=0$, which is equivalent to the condition~\eqref{eq:Jc-Neven}. In the ground state characterized by $s = -N/2$ and $d = 0$, the source TLSs are perfectly aligned, while in the ground state with $s = 0$ and $d = N/2$, they are perfectly antialigned. Therefore, we refer to these ground states as ferromagnetic and antiferromagnetic, respectively.

\subsection{Effect of \texorpdfstring{$N$}{{\it N}}} 
For $T>0$, $C(T)\to 1$ as $N\to\infty$, which is similar to the $J>0$ case discussed in Sec.~\ref{sec:C-N-ferro}, for mathematical details see App.~\ref{sec:appendix-D-large-N}. At $T=0$, for the antiferromagnetic ground state, the coherence is zero, leading to discontinuity of $C$ at $T=0$, while for the ferromagnetic ground state it converges to one, see Eqs.~\eqref{eq:C0-Jnegative-Neven} and~\eqref{eq:C0-result}. The position of the level-crossing line~\eqref{eq:Jc-Neven} in the phase diagram of Fig.~\ref{fig:fig3_negJ_evenN}~a) is influenced by $N$ weakly (compared to other parameters). If $\omega_0/\gamma N\ll 1$, the line is given by $J_{\rm lc}+\omega_{\rm a,lc}/2+\gamma_{\rm lc}/2 = 0$.

\section{Temperature dependence of \texorpdfstring{$C$}{{\it C}} for antiferromagnetic interaction \texorpdfstring{($J<0$)}{ } and odd \texorpdfstring{$N$}{{\it N}}}
\label{sec:C-antiferromagnetic-odd}

Finally, we consider the case where $J<0$ and $N$ is odd. The main differences from the even-$N$ case lie in the nonzero value of $C$ near $T=0$ for $J<J_{\rm lc}$, as well as a more complex temperature dependence of $C$. All quantitative results for $N$ odd are derived analogously to those for $N$ even. Therefore, we focus primarily on the qualitatively new features that arise in comparison to the even-$N$ case.

\subsection{Ground state energy level crossing}
 
For fixed $s$ the minimal energy $E^{\rm min}_s$ is given by Eq.~\eqref{eq:minimumE}. Because $E^{\rm min}_{|s|}>E^{\rm min}_{-|s|}$, the minimum is reached for $s_{\rm min}<0$.
Analyzing the derivative $\partial E^{\rm min}_{s}/\partial s$ for $s\in (-N/2,-1/2)$ leads to the conclusion that the minimum energy cannot be reached for such $s$. Thus, minima can occur only at the boundary values $s_{\rm min}=-1/2$ and $s_{\rm min}=-N/2$. Introducing the energy difference 
\begin{equation} 
\Delta E^{\rm min} = E^{\rm min}_{-1/2}-E^{\rm min}_{-N/2}, 
\end{equation} 
the level crossing occurs at $\Delta E^{\rm min}=0$, which is equivalent to the condition:
\begin{equation}{\label{eq:Jcritodd}}
2(N_{\rm lc}-1) J_{\rm lc}= \sqrt{\gamma_{\rm lc}^2+\omega_{0,{\rm lc}}^2}-\sqrt{\gamma_{\rm lc}^2N_{\rm lc}^2+\omega_{0,{\rm lc}}^2}+\omega_{\rm a,lc}-N_{\rm lc}\omega_{\rm a,lc}.
\end{equation}
For $N\to\infty$, this equation simplifies to
\begin{equation}
J_{\rm lc}=-\frac{\gamma_{\rm lc}+\omega_{\rm a,lc}}{2}.
\end{equation}
In the antiferromagnetic phase, we have $E^{\rm min}_{-1/2}<E^{\rm min}_{-N/2}$, hence $s_{\rm min}=-1/2$, and 
\begin{equation}{\label{eq:C0-result-oddN}}
C_0=\frac{\gamma}{\sqrt{\gamma^2+\omega_0^2}}.
\end{equation}
In the ferromagnetic phase, $E^{\rm min}_{-1/2}>E^{\rm min}_{-N/2}$, thus $s_{\rm min}=-N/2$, and we get 
\begin{equation}
C_0=\frac{\gamma N}{\sqrt{\gamma^2N^2+\omega_0^2}}.
\end{equation}

\subsection{Effect of \texorpdfstring{$J$}{\textit{J}}, \texorpdfstring{$\omega_{\rm a}$}{omega{\scriptsize a}}, and \texorpdfstring{$\gamma$}{gamma}}

In the antiferromagnetic phase, the key difference from the even-$N$ case is the shape of $C(T)$. The value of this function at $T=0$ is generally neither a maximum nor a minimum. While a plateau still appears near $T=0$, it may be followed by one of several behaviors: an increase to a maximum followed by a monotonic decline; a decrease, then an increase to a local maximum, and subsequently a monotonic decline; a purely monotonic decline. In the second scenario, the maximum reached after the initial dip can be either greater or smaller than the coherence value at $T=0$. Similar features are observed when varying $\omega_{\rm a}$ and $\gamma$.

\section{Summary and perspectives} 
The interaction within the environment, i.e., between the source two-level systems, can have a significant impact on the quantum coherence induced by these TLSs in a target TLS. The presence of such an interaction does not inherently suppress the coherence, thereby affirming the robustness and potential applicability of the reported phenomenon in quantum systems strongly coupled to their surrounding environment. 

To describe the effects caused by the intersource interaction, we have derived the density operator of the target TLS and an exact analytical expression for the autonomous coherence $C$ generated in the target TLS, assuming a thermal state for the entire system. The coherence has been analyzed as a function of the intersource coupling constant $J$, the ambient temperature $T$, and other model parameters: the source TLSs gap $\omega_{\rm a}$, the target-source coupling constant $\gamma$, and the number $N$ of source TLSs. 

For $T>0$, we have proven that the coherence is an increasing function of  $J$,  $\omega_{\rm a}$, and $\gamma$, i.e., that the inequalities $C(T,J_1)<C(T,J_2)$, $C(T,\omega_{{\rm a}1})<C(T,\omega_{{\rm a}2})$, and $C(T,\gamma_1)<C(T,\gamma_2)$ hold for $J_1<J_2$, $\omega_{{\rm a}1} < \omega_{{\rm a}2}$, and $\gamma_1<\gamma_2$, respectively.

An upper bound on $C$ as a function of $T$ has been derived in the limit $J\to\infty$. This bound corresponds to the coherence generated by a single source TLS with the energy gap $N\omega_{\rm a}$ interacting with the target TLS via a renormalized coupling $\gamma N$. The bound is given by a nontrivial yet analytically tractable function, whose explicit form enables qualitative insights into the influence of model parameters on $C$.

Furthermore, $C$ vanishes in the high-temperature ($T\to\infty$), weak-coupling ($\gamma\to0$) and the small-gap limit ($\omega_{\rm a}\to0$). Although the coherence is generated via the interaction with the source TLSs, the maximal value $C=1$ is not achieved in either the strong-coupling ($\gamma \to \infty$) or the large-gap limit ($\omega_{\rm a} \to \infty$) when $T>0$. The value of $C = 1$ can be approached asymptotically in the thermodynamic ($N\to\infty$) and the low-temperature limit ($T\to 0$).

The functional form of $C(T)$ is influenced by a change in the ground state of the system, which occurs for $T=0$ and $J<0$. We distinguish a ferromagnetic ground state in which all source TLSs are localized at lower energy levels and an antiferromagnetic ground state characterized by an alternating configuration of source TLSs in their ground and excited states. The analysis of $C(T)$ has been carried out separately for the two ground states. Additionally, the cases of even and odd $N$ differ when the ground state is antiferromagnetic.  

When the ground state is ferromagnetic, e.g.\ when $J>0$, the coherence $C(T)$ is a monotonically decreasing function of $T$, exhibiting a plateau in the vicinity of $T=0$. The maximum coherence $C(0)$, attained at $T=0$, depends on $N$, $\gamma$, and $\omega_0$. These parameters can be adjusted to make $C(0)$ arbitrarily close to 1. An increase in $\gamma$ or $N$ leads to a global enhancement of the coherence for all values of $T$. In contrast, increasing $J$ or $\omega_{\rm a}$ enhances the coherence away from the plateau, but leaves $C(0)$ unaffected. 

For the antiferromagnetic ground state and even $N$, the coherence vanishes at $T=0$ and it remains vanishingly small within a plateau in the vicinity of $T=0$. Beyond this region, $C(T)$ rises to a pronounced maximum at some finite $T$, followed by the decay toward zero in the high-temperature limit. If $N$ is odd, $C(0)$ is small compared to the ferromagnetic case, but remains finite and depends solely on $\gamma$ and $\omega_0$. Consequently, the temperature dependence of $C$ is richer: the plateau near $T=0$ may be followed by an initial increase or decrease of $C(T)$.

The general aim of investigations on the autonomous coherence is the development of reliable and reusable schemes for producing the coherence without external time-dependent driving. Our exact results and corresponding asymptotic analysis have revealed basic properties of the autonomously generated local quantum coherence and strategies for enhancing this coherence at finite temperatures. The insights gained can be relevant for the following directions (i)-(iii) of further research. 

(i) The present results may serve as a starting point for grasping coherence behavior in more general models of the target TLS environment. This includes, for example, nonintegrable models and experimentally relevant spin systems with anisotropic Heisenberg interactions, where no exact solution is available. 

(ii) Further generalizations of the model toward applicability may include finite-time interactions between the target TLS and the environment, as well as explicit measurement processes. A promising direction is the incorporation of a modified input-output formalism tailored to quantum pulses \cite{Kiilerich/etal:2019, Kiilerich/etal:2020}, which may enable the development of a practical method for characterizing the autonomous coherence under realistic experimental conditions.

The studies outlined in (i) and (ii) are essential for identifying systems and parameter regimes that are most suitable for an experimental realization of autonomously generated local quantum coherence in a target TLS. Although the specific physical platform has yet to be established, initial experimental investigations of relevant interaction mechanisms, such as those arising from electron-nuclear spin coupling in quantum dots~\cite{Atature/etal:2018, Appel/etal:2025}, nitrogen-vacancy centers in diamond~\cite{Du/etal:2024}, or coupling schemes among superconducting qubits~\cite{Blais/etal:2004, Koch/etal:2007, Forn-Diaz/etal:2010, Hamedani/etal:2021, Pekola/Karimi:2021, Guthrie/etal:2022}, may already be within experimental reach. 

For superconducting qubits, a transmon qubit serving as the target system can be coupled to $N$ auxiliary (source) transmons or low-population resonators, extending the experimental setup of Ref.~\cite{Guthrie/etal:2022}. In this setup, the qubit-oscillator coupling approaches the required qubit-qubit coupling at low occupation. Typical transmon transition frequencies are $\omega_0/2\pi\in (1,10)$ GHz, and coupling constants $\gamma/2\pi$, $J/2\pi\in (0.1,1)$~GHz \cite{Forn-Diaz/etal:2010, Guthrie/etal:2022}. The experiments are performed in dilution refrigerators at temperatures of $T\in (10,100)$~mK \cite{Guthrie/etal:2022}. The refrigerator environment can act as a weakly coupled bath that prepares a global Gibbs state. Furthermore, capacitive, inductive, or SQUID couplers can mediate interactions between the source qubits, using techniques reported in Refs.~\cite{Hamedani/etal:2021, Guo/etal:2021}. Proof-of-principle tests of the predicted effects require only a relatively short Ising chain with $N < 10$, which future multiqubit processors may be able to access \cite{Guo/etal:2021}. The target-qubit coherence can be determined using standard single-qubit quantum state tomography \cite{Bianchetti/etal:2010}. For the parameter estimates considered here, this procedure is expected to yield $C(T)$ up to 0.994. Larger values of $N$, and thus larger $C(T)$, may be attainable in systems of hyperfine-coupled quantum dots \cite{Appel/etal:2025, Khaetskii/etal:2002, Merkulov/etal:2002, Coish/Loss:2004, Urbaszek/etal:2013, Zaporski/etal:2023}.

(iii) Simultaneously, it is worth to explore the role of the autonomous local coherence in quantum thermodynamic processes, for example, through its influence on entropy, entropy production, and ergotropy \cite{santos:2019, Zicari/etal:2023, Rodrigues/Lutz:2024}. The primary objective of such investigations is to assess the potential of the local coherence to enhance the performance of quantum machines~\cite{Cangemi/etal:2024, Huang/etal:2025} operating under strong coupling conditions~\cite{Newman/etal:2017, Giannelli/etal:2024}.

\section*{Acknowledgments}

Y.H.\ and A.R.\ gratefully acknowledge financial support by the Czech Science Foundation (Project No.\ 26-22148S). M.K.\ and R.F.\ acknowledge the grant 22-27431S of the Czech Science Foundation. M.K.\ also acknowledges the Programme Johannes Amos Comenius under the Ministry of Education, Youth and Sports (MEYS) of the Czech Republic reg.\ no.\ CZ.02.01.01/00/22\_008/0004649 (QUEENTEC). A.R.\ furthermore acknowledges financial support from Charles University via the project PRIMUS/22/SCI/009.

\appendix 

\section{Factorized eigenvectors of the total Hamiltonian}
\label{sec:appendixA}

The interaction of the target TLS and the source TLSs is described by the separable Hamiltonian of the form 
\begin{equation}
\label{eq:Hint_separable}
\frac{\gamma}{2} \sum_{i=1}^N \hat\sigma_0^x \hat\sigma_i^z=\frac{\gamma}{2}\hat{\sigma}^x_0  \hat{S}_{\rm a},
\end{equation} 
see Eqs.~\eqref{eq:Sa-def} and~\eqref{eq:Htot-Ising-S}. Moreover, the commutation relation 
\begin{equation}
\label{eq:commutator}
\left[\frac{\omega_{\rm a}}{2} \hat{S}_{\rm a} -\frac{J}{2}\sum_{i=1}^{N}\hat{\sigma}^z_i\hat{\sigma}^z_{i+1},\hat{S}_{\rm a}\right] =0
\end{equation} 
holds since  $\hat{S}_{\rm a}$ and $\hat{\sigma}_i^z$ commute for all $i \in \{1, \dots, N\}$.
Equations~\eqref{eq:Hint_separable} and~\eqref{eq:commutator} imply that eigenvectors $\ket{\tau}$ of the total Hamiltonian $\hat{H}$, given in Eq.~\eqref{eq:Htot-Ising-S}, can be written in the factorized form 
\begin{equation}
\ket{\tau}=\ket{\varphi}\ket{\psi},
\end{equation} 
where $\ket{\varphi}$ is a vector in the Hilbert space of the target TLS and $\ket{\psi}$ of the source TLSs. 

There exists a general class of target-source interactions for which such a factorization holds true. To see this, let us consider systems A and B described by $\hat{H}_{\rm A}$ and $\hat{H}_{\rm B}$ having finite-dimensional Hilbert spaces $\mathcal{H}_{\rm A}$ and $\mathcal{H}_{\rm B}$. These systems interact via the  Hamiltonian $H^{\rm AB}_{\rm int}$, yielding the total Hamiltonian of the form
\begin{equation}
\hat{H}=\hat{H}_{\rm A}+\hat{H}^{\rm AB}_{\rm int}+\hat{H}_{\rm B}.
\end{equation} 
We assume 
\begin{align}
&\label{eq:gen-assumption1}\hat{H}^{\rm AB}_{\rm int}=\hat{H}^{\rm A}_{\rm int}\otimes\hat{H}^{\rm B}_{\rm int},\\
&\label{eq:gen-assumption2}[\hat{H}^{\rm B}_{\rm int},\hat{H}_{\rm B}]=0,
\end{align}
i.e., that $\hat{H}^{\rm AB}_{\rm int}$ is separable and its projection $\hat{H}^{\rm B}_{\rm int}$ on $\hat{H}_{\rm B}$ commutes with $\hat{H}_{\rm B}$. As a consequence, there exist common eigenvectors $\ket{\psi_{j}}$, $j\in\{1,\dots,\rm{dim}\;\mathcal{H}_{\rm B}\}$ of $\hat{H}_{\rm B}$ and $\hat{H}^{\rm B}_{\rm int}$ that form a basis in $\mathcal{H}_{\rm B}$. $\hat{H}_{\rm B}$ and $\hat{H}^{\rm B}_{\rm int}$ act on these eigenvectors $\ket{\psi_{j}}$ as follows:
\begin{align}
&\hat{H}_{\rm B}\ket{\psi_{j}}=\lambda_{j}\ket{\psi_{j}},\\
&\hat{H}^{\rm B}_{\rm int}\ket{\psi_{j}}=\mu_{j}\ket{\psi_{j}},
\end{align} 
where $\lambda_{j}$ and $\mu_{j}$ are real since the Hamiltonians are Hermitian operators.

Let $\ket{\varphi} \in \mathcal{H}_{\rm A}$, then $\hat{H}$ acts on $\ket{\varphi}\ket{\psi_{j}}$ as 
\begin{align}
\hat{H}\ket{\varphi}\ket{\psi_{j}}=\left((\hat{H}_{\rm A}+\mu_{j}\hat{H}^{\rm A}_{\rm int}+\lambda_{j}\hat{I})\ket{\varphi}\right)\ket{\psi_{j}}.
\end{align}
For each $j$, $(\hat{H}_{\rm  A}+\mu_{j}\hat{H}^{\rm A}_{\rm int}+\lambda_{j}\hat{I})$ is a Hermitian operator, and its eigenvectors $\ket{\varphi_{j,i}}$ to the eigenvalues $E_{j,i}$, $i\in\{1,\dots,\rm{dim}\;\mathcal{H}_{\rm A}\}$, form a basis of $\mathcal{H}_{\rm A}$.
Moreover, $\ket{\varphi_{j,i}}$ are eigenvectors of $\hat{H}_{\rm A}+\mu_{j}\hat{H}^{\rm A}_{\rm int}$, since every vector in $\mathcal{H}_{\rm A}$ is an eigenvector of $\lambda_{j}\hat{I}$. The ket $\ket{\varphi_{j,i}}\ket{\psi_{j}}$ is an eigenvector of $\hat{H}$ with eigenvalue $E_{j,i}$. Kets $\ket{\varphi_{j,i}}\ket{\psi_{j}}$ are orthogonal due to the orthogonality of $\ket{\psi_j}$, 
\begin{equation}
\left(\bra{\psi_{j}}\bra{\varphi_{j,i}}\right) \left(\ket{\varphi_{j',i'}} \ket{\psi_{j'}}\right) = \bra{\psi_{j}}\ket{\psi_{j'}} \bra{\varphi_{j,i}}\ket{\varphi_{j',i'}}=\delta_{j,j'}\delta_{i,i'},
\end{equation} 
where $\delta_{i,j}$ is the Kronecker delta. 

By the Steinitz exchange lemma, applicable to finite-dimensional spaces, the kets 
$\ket{\varphi_{j,i}}\ket{\psi_j}$, for 
$i\in\{1,\dots,\rm{dim}\;\mathcal{H}_{\rm A}\}$ and 
$j\in\{1,\dots,\rm{dim}\;\mathcal{H}_{\rm B}\}$, form a basis of the whole Hilbert space of the system (A+B), and each of these kets is an eigenvector of $\hat{H}$.

For the interaction Hamiltonian~\eqref{eq:Hint-Ising}, the conditions~\eqref{eq:gen-assumption1} and~\eqref{eq:gen-assumption2} are satisfied because Eqs.~\eqref{eq:Hint_separable} and~\eqref{eq:commutator} hold.

\section{Proofs that \texorpdfstring{$C$}{\textit{C}} is an increasing function of \texorpdfstring{$J$}{\textit{J}}, \texorpdfstring{$\omega_{\rm a}$}{omega{\scriptsize a}}, and \texorpdfstring{$\gamma$}{gamma}} 
\label{sec:appendix-C-monotonic-all}

We derive the inequalities~\eqref{eq:C-inequality-J}, \eqref{eq:C-inequality-wa}, and~\eqref{eq:C-inequality-gamma} by showing that $C$ given in Eq.~\eqref{eq:C-R} respectively satisfies 
$\partial C / \partial J >0$, $\partial C / \partial \omega_{\rm a} >0$, and $\partial C / \partial \gamma >0$ for any $J$, $\omega_{\rm a}$, and $\gamma$. Clarity of the derivations benefits from using the short-hand notation 
\begin{equation} 
\label{eq:C-productform}
C=\frac{\displaystyle\sum_{i>0}r_i\, l_i^{(\gamma)} s_i^{(\omega_{\rm a})} s_i^{(\gamma)}}{\displaystyle \sum_{i\geq0}r_i\, c_i^{(\omega_{\rm a})} c_i^{(\gamma)}},
\end{equation} 
where $r_i = e^{\beta JN/2} R(i,N)$, are rescaled weighting factors with $R(i,N)$ defined in Eq.~\eqref{eq:R}, hence 
\begin{equation}
\label{eq:ri-def} 
r_i=\sum_{d=0}^{\lfloor N/2 \rfloor} \Omega(i,d,N)\, e^{\beta J(N-2d)}. 
\end{equation}
After the rescaling by $e^{\beta JN/2} $, all the exponents $\beta J (N-2d)$ have the same sign as $J$ since $d\leq N/2$. Note that $\Omega(i,d,N)=0$ if $d>N/2-i$, see the definition~\eqref{eq:Omega-result}. Furthermore, we have introduced 
\begin{subequations}
\label{eq:ci-si} 
\begin{align} 
&\label{eq:l-gamma} l_i^{(\gamma)} =\frac{2\gamma i}{\sqrt{\omega_0^2+(2\gamma i)^2}},\\ 
& s_i^{(\omega_{\rm a})}=\sinh(\beta\omega_{\rm a} i),\\
& c_i^{(\omega_{\rm a})}  =\cosh(\beta \omega_{\rm a} i),\\[1ex] 
&\label{eq:si-gamma} s_i^{(\gamma)} =\sinh(\frac{\beta}{2} \sqrt{\omega_0^2 + (2\gamma i)^2}), \\[1ex]
& c_0^{(\gamma)} =\begin{cases} \frac{1}{2}\cosh\left(\frac{\beta\omega_0}{2}\right) & \textrm{for}\ N\ \textrm{even}, \\ 0 &\textrm{for}\ N\ \textrm{odd}, \end{cases} \\[1ex]
&\label{eq:ci-gamma} c_i^{(\gamma)}= \cosh(\frac{\beta}{2} \sqrt{\omega_0^2 + (2\gamma i)^2}), \quad i> 0. 
\end{align}
\end{subequations}
This notation explicitly distinguishes functions of $\omega_{\rm a}$ and $\gamma$, whereas $J$ appears only in~$r_i$.

\subsection{\texorpdfstring{$\partial C/\partial J >0$}{{\it C} is a monotone function of {\it J}}}
\label{sec:appendix-C-monotonic}
The summands in Eq.~\eqref{eq:C-productform} are factorized according to their dependence on $J$, $\omega_{\rm a}$, and $\gamma$. For showing that $\partial C/\partial J >0$ holds for any $J$, we group functions of $\omega_{\rm a}$ and $\gamma$ and 
$s_i=l_i^{(\gamma)} s_i^{(\omega_{\rm a})} s_i^{(\gamma)}$,
$c_i=c_i^{(\omega_{\rm a})} c_i^{(\gamma)}$, and introduce the derivative $r'_i$ defined by  
\begin{equation} 
\label{eq:ri-prime-def}
r'_i=\frac{\partial r_i}{\partial J} = 
\beta \sum_{d=0}^{\lfloor N/2 \rfloor}(N-2d) \Omega(i,d,N)\, e^{\beta J(N-2d)} . 
\end{equation} 
In Eqs.~\eqref{eq:ri-def} and~\eqref{eq:ri-prime-def}, the summation range is formally extended up to $\lfloor N/2 \rfloor$. Several terms in such a sum can be zero, since the degeneracy $\Omega(i,d,N)$ vanishes if $d>N/2-i$, see its definition~\eqref{eq:Omega-result}.

The sign of  
\begin{equation}
\label{eq:C-partial-J}
\frac{\partial C}{\partial J}=\frac{\displaystyle\sum_{i>0}r_i' s_i\displaystyle\sum_{j\geq0}r_j c_j-\displaystyle\sum_{i>0}r_i s_i\displaystyle\sum_{j\geq0}r_j' c_j}{\left( \displaystyle\sum_{i\geq0}r_i c_i\right)^2}
\end{equation}
is determined by the numerator, where we may reorder the summands as follows 
\begin{align} 
\label{eq:dCdJ-numerator}
\displaystyle\sum_{i>0,j\geq0} (r_i' r_j- r_i r_j')s_ic_j 
=\sum_{0<j<i}(r'_ir_j-r_ir_j')(s_ic_j-s_jc_i)
+\sum_{i>0}(r'_ir_0-r_ir_0')s_ic_0. 
\end{align}
We will now show that the right-hand side of Eq.~\eqref{eq:dCdJ-numerator} is positive. 

By definition, see Eqs.~\eqref{eq:ci-si}, $s_ic_0>0$ for $N$ even and $s_ic_0=0$ for $N$ odd. 
Also, we have $(s_ic_j-s_jc_i) = c_ic_j(s_i/c_i-s_j/c_j) >0$ for all $0<j<i$, which follows from the fact that the ratio 
\begin{equation}
\frac{s_i}{c_i} = \frac{2\gamma i }{\sqrt{\omega_0^2+(2\gamma i)^2}}\tanh(\beta \omega_{\rm a} i)\tanh(\frac{\beta}{2} \sqrt{\omega_0^2 + (2\gamma i)^2}), 
\end{equation}
is a monotone increasing function of $i$. 

Therefore, for $\partial C/\partial J>0$ to hold, it remains to be demonstrated that $(r'_ir_j-r_ir_j')>0$ for $0\leq j<i$. 
To see this, we again rearrange corresponding summands such that the summation indexes in double sums are ordered 
\begin{align} 
\label{eq:rirj-difference}
&\frac{r'_ir_j-r_ir_j'}{2\beta}= \sum_{d,f=0}^{\lfloor N/2 \rfloor}(f-d)\Omega(i,d,N)\Omega(j,f,N)\,e^{\beta J(2N-2d-2f)}\nonumber\\
&\hspace{5ex} =\sum_{d<f}(f-d)\Big(\Omega(i,d,N)\Omega(j,f,N)-\Omega(i,f,N)\Omega(j,d,N)\Big) e^{\beta J(2N-2d-2f)}\, ,
\end{align}
and verify the inequality 
\begin{equation}
\label{eq:Omega-inequality}
\Omega(i,d,N)\Omega(j,f,N)-\Omega(i,f,N)\Omega(j,d,N)\geq0,
\end{equation}
for all $i,j, d$, and $f$ within the entire summation range, i.e., for  
$0<j<i\leq N/2$, $0\leq d<f\leq \lfloor N/2 \rfloor$. 

In Eq.~\eqref{eq:rirj-difference}, there exist summands with such  $i,j, d$, and $f$ that at least one of the degeneracies $\Omega$ vanishes. 
The negative term in~\eqref{eq:Omega-inequality} becomes zero 
\begin{itemize} 
\item[(A)] for $i=N/2$ and $f>0$, where $\Omega(i,f,N)=0$; 
\item[(B)] for $d=0$ and $j< N/2$, where $\Omega(j,d,N)=0$. 
\end{itemize}
If the positive term in~\eqref{eq:Omega-inequality} vanishes, the negative one vanishes as well: 
\begin{itemize}  
\item[(C)] for $d>N/2-i$, $\Omega(i,d,N)=0$, and since $f>d>N/2-i$, we have $\Omega(i,f,N)=0$;
\item[(D)] for $f>N/2-j$, $\Omega(j,f,N)=0$, and due to $f>N/2-j>N/2-i$, $\Omega(i,f,N)=0$. 
\end{itemize}
This shows that in all cases (A)-(D) the inequality~\eqref{eq:Omega-inequality} holds true. 
For all other considered $i,j, d$, and $f$, both terms on the left-hand side of~\eqref{eq:Omega-inequality} are nonzero. We rewrite 
$\Omega(i,d,N)\Omega(j,f,N)-\Omega(i,f,N)\Omega(j,d,N) = 
\Omega(i,f,N)\Omega(j,f,N)\Big( \Omega(i,d,N)/\Omega(i,f,N)-\Omega(j,d,N)/\Omega(j,f,N)\Big)$ and check that 
\begin{equation}
\label{eq:Omega-Omega-inequality}
\frac{\Omega(i,d,N)}{\Omega(i,f,N)}> \frac{\Omega(j,d,N)}{\Omega(j,f,N)} 
\end{equation}
holds. 
It is sufficient to verify~\eqref{eq:Omega-Omega-inequality} for $i=j+1$:
\begin{align} 
    \frac{\frac{\Omega(j+1,d,N)}{\Omega(j+1,f,N)}}{\frac{\Omega(j,d,N)}{\Omega(j,f,N)}}
    =\frac{(N/2-j-d)(N/2+j-f)}{(N/2-j-f)(N/2+j-d)}
    =\frac{A+j(f-d)}{A-j(f-d)}>1
\end{align}
where $A=N^2/4+df-j^2-(d+f)N/2$ and $d<f$. 
This concludes the proof of~\eqref{eq:Omega-inequality} and hence also the proof that $\partial C / \partial J >0$.

\subsection{\texorpdfstring{$\partial C/\partial \omega_{\rm a} >0$}{{\it C} is a monotone function of {\it J}}}
\label{sec:appendix-C-monotonic_omega_a}
To see that $\partial C/\partial\omega_{\rm a}>0$ we start with gathering functions in Eq.~\eqref{eq:C-productform}: $a_i=l_i^{(\gamma)} s_i^{(\gamma)}$,
$b_i=c_i^{(\gamma)}$, $s_i=s_i^{(\omega_{\rm a})}$, $c_i=c_i^{(\omega_{\rm a})}$. The derivatives of $\omega_{\rm a}$-containing functions satisfy
$\partial c_i/ \partial\omega_{\rm a}= 
\beta is_i$, 
$\partial s_i/\partial\omega_{\rm a}= \beta ic_i$, 
hence we have
\begin{equation} 
\label{eq:C-partial-omega_a}
\frac{1}{\beta}\frac{\partial C}{\partial \omega_{\rm a}}= 
\frac{\displaystyle\sum_{i>0,j\geq0}ir_ia_ic_i  r_jb_jc_j-\displaystyle\sum_{i>0,j>0}j r_ia_is_i r_jb_js_j}{\left( \displaystyle\sum_{i\geq0}r_ib_i c_i\right)^2}.
\end{equation} 
Next, we reorder summands in the numerator as follows   
\begin{align} 
\label{eq:dCdomega_a-numerator}
& \displaystyle\sum_{ i,j>0}r_ir_j a_ib_j(i c_ic_j-j s_i s_j)+\sum_{i>0}ir_ia_ic_i r_0b_0c_0\\
& =\sum_{0<i<j}r_ir_j\big(c_ic_j(ia_ib_j+ja_jb_i)-s_is_j(ia_jb_i+ja_ib_j) \big)+ \sum_{i>0}ir_i^2a_ib_i+\sum_{i>0}ir_ia_ic_i r_0b_0c_0,
\nonumber 
\end{align} 
where the second term on the right-hand accounts for all terms with $i=j$, where we have  
$(i c_ic_j-j s_i s_j) = c_i^2-s_i^2=1$.

To prove the inequality $\partial C / \partial \omega_{\rm a} >0$, we show that the right-hand side of Eq.~\eqref{eq:dCdomega_a-numerator} is positive:
The second summand $\sum_{i>0}ir_i^2a_ib_i$ on the right-hand side of Eq.~\eqref{eq:dCdomega_a-numerator} is positive; the third one vanishes for $N$ odd and is positive for $N$ even. 
It remains to demonstrate that $c_ic_j(ia_ib_j+ja_jb_i)>s_is_j(ia_jb_i+ja_ib_j)$ for $0<i<j$. Since the both sides of the last yet-to-be-proven inequality are nonzero, we can instead consider the ratio  
$c_ic_j(ia_ib_j+ja_jb_i)/s_is_j(ia_jb_i+ja_ib_j)$, 
which is equal to 
$c_ic_j(ia_i/b_i +j a_j/b_j)/s_is_j(ia_j/b_j+ja_i/b_i)$ and it satisfies the inequality 
\begin{equation}
\label{eq:ratio-inequality-omegaa}
\frac{c_i}{s_i}\frac{c_j}{s_j}\frac{i\frac{a_i}{b_i}+j\frac{a_j}{b_j}}{i\frac{a_j}{b_j}+j\frac{a_i}{b_i}}>1,
\end{equation}
because 
$c_l/s_l=\coth(\beta \omega_{\rm a} l)>1$, the ratio
\begin{equation} 
\frac{a_i}{b_i}=\frac{2\gamma i }{\sqrt{\omega_0^2+(2\gamma i)^2}}\tanh(\frac{\beta}{2} \sqrt{\omega_0^2 + (2\gamma i)^2}),
\end{equation}
obeys $a_i/b_i<a_j/b_j$ for $i<j$ and also   
\begin{equation}
\frac{i\frac{a_i}{b_i}+j\frac{a_j}{b_j}}{i\frac{a_j}{b_j}+j\frac{a_i}{b_i}}>1
\end{equation}
holds, which is an example of the rearrangement inequality \cite{Hardy/etal:1952}. 
This completes the proof of~\eqref{eq:ratio-inequality-omegaa}. Thus, it indeed holds that the numerator in Eq.~\eqref{eq:C-partial-omega_a} is positive, which concludes the proof of inequality $\partial C / \partial \omega_{\rm a} >0$.

\subsection{\texorpdfstring{$\partial C/\partial \gamma >0$}{{\it C} is a monotone function of {\it J}}}
\label{sec:appendix-C-monotonic_gamma}
We now show that $\partial C / \partial \gamma >0$ for $\gamma\in (0,\infty)$ holds. For notational convenience, we rename functions from Eq.~\eqref{eq:C-productform}: $k_i=l_i^{(\gamma)}/\beta,c_i=c_i^{(\gamma)},s_i=s_i^{(\gamma)},a_i=s_i^{(\omega_{\rm a})}$, and $b_i=c_i^{(\omega_{\rm a})}$. Derivatives of $\gamma-$containing functions satisfy: $k_i'=(1/\beta)(\partial l_i/\partial \gamma)= \beta^{-1}(2 i \omega_0^2)/(4\gamma^2i^2+\omega_0^2)^{3/2}>0 $, $\partial c_i/\partial \gamma=\beta i l_is_i$ and $\partial s_i/\partial \gamma=\beta i l_ic_i$ hence we have
\begin{equation} 
\label{eq:C-partial-gamma}
\frac{1}{\beta}\frac{\partial C}{\partial \gamma}= 
\frac{\displaystyle\sum_{i>0,j\geq0}k_i'r_ia_is_i  r_jb_jc_j+\displaystyle\sum_{i>0,j\geq0}ik_ir_ia_ic_i  r_jb_jc_j-\displaystyle\sum_{i>0,j>0}j r_ia_is_i k_jr_jb_js_j}{\left( \displaystyle\sum_{i\geq0}r_ib_i c_i\right)^2}.
\end{equation} 
Next, we reorder summands in the numerator as follows   
\begin{align}
\label{eq:dCdgamma-numerator}& \displaystyle\sum_{i,j>0}k_ir_ir_j a_ib_j(i k_ic_ic_j-j k_js_i s_j)+\sum_{i>0}ik_ir_ia_ic_i r_0b_0c_0+\displaystyle\sum_{i>0,j\geq0}k_i'r_ia_is_i  r_jb_jc_j\\
& =\sum_{0<i<j}r_ir_j\big(c_ic_j(ik_i^2a_ib_j+jk_j^2a_jb_i)-s_is_j(ik_ik_ja_jb_i+jk_ik_ja_ib_j) \big)\nonumber\\&+ \sum_{i>0}ir_i^2k_i^2a_ib_i+\sum_{i>0}ik_ir_ia_ic_i r_0b_0c_0+\displaystyle\sum_{i>0,j\geq0}k_i'r_ia_is_i  r_jb_jc_j,
\nonumber 
\end{align} 
where the second term on the right-hand accounts for all terms with $i=j$, where we have  
$(ik_i c_ic_j-jk_j s_i s_j) =ik_i( c_i^2-s_i^2)=ik_i$.

To prove the inequality $\partial C / \partial \gamma >0$, we show that the right-hand side of Eq.~\eqref{eq:dCdgamma-numerator} is positive:
The second summand $\sum_{i>0}ir_i^2k_i^2a_ib_i$ and the fourth summand $\displaystyle\sum_{i>0,j\geq0}k_i'r_ia_is_i  r_jb_jc_j$ on the right-hand side of Eq.~\eqref{eq:dCdgamma-numerator} are positive; the third one vanishes for $N$ odd and is positive for $N$ even. 
It remains to demonstrate that $c_ic_j(ik_i^2a_ib_j+jk_j^2a_jb_i)>s_is_jk_ik_j(ia_jb_i+ja_ib_j)$ for $0<i<j$. Similarly to the previous proof, we can consider the ratio  
$c_ic_j(ik_i^2a_ib_j+jk_j^2a_jb_i)/s_is_jk_ik_j(ia_jb_i+ja_ib_j)$, 
which is equal to 
$c_ic_j(ik_i^2a_i/b_i +jk_j^2 a_j/b_j)/s_is_j(ik_ik_ja_j/b_j+jk_ik_ja_i/b_i)$ and it satisfies the inequality 
\begin{equation}
\label{eq:ratio-inequality-gamma}
\frac{c_i}{s_i}\frac{c_j}{s_j}\frac{(ik_i)(k_i\frac{a_i}{b_i})+(jk_j)(k_j\frac{a_j}{b_j})}{(ik_i)(k_j\frac{a_j}{b_j})+(jk_j)(k_i\frac{a_i}{b_i})}>1,
\end{equation}
because $c_l/s_l=\coth(\left. \beta \sqrt{\omega_0^2+(2\gamma l)^2} \right/2)>1$, the ratios
\begin{align} 
&\frac{k_ia_i}{b_i}=\frac{2\gamma i }{\sqrt{\omega_0^2+(2\gamma i)^2}}\tanh(\beta i\omega_{\rm a}),\\
&ik_i=\frac{2\gamma i^2 }{\sqrt{\omega_0^2+(2\gamma i)^2}}
\end{align}
obey $k_ia_i/b_i<k_ja_j/b_j$ and $ik_i<jk_j$ for $i<j$ and also   
\begin{equation}
\frac{(ik_i)(k_i\frac{a_i}{b_i})+(jk_j)(k_j\frac{a_j}{b_j})}{(ik_i)(k_j\frac{a_j}{b_j})+(jk_j)(k_i\frac{a_i}{b_i})}>1
\end{equation}
holds by rearrangement inequality \cite{Hardy/etal:1952}. 
This completes the proof of~\eqref{eq:ratio-inequality-gamma}. Consequently, the numerator in Eq.~\eqref{eq:C-partial-gamma} is positive, thereby proving that $\partial C / \partial \gamma > 0$.

\section{Large-\texorpdfstring{$N$}{{\it N}} asymptotic behavior of \texorpdfstring{$C(T)$}{{\it C(T)}}
\label{sec:appendix-D-large-N}}

Coherence $C_0$ at $T=0$ is given by Eq.~\eqref{eq:C0-result} for the ferromagnetic ground state, and by Eqs.~\eqref{eq:C0-Jnegative-Neven} and~\eqref{eq:C0-result-oddN} for the antiferromagnetic ground state for $N$ even and odd, respectively. Parameter regions, where either ferromagnetic or antiferromagnetic ground state is realized, are separated by Eq.~\eqref{eq:Jc-Neven} for $N$ even and by~\eqref{eq:Jcritodd} for $N$ odd. Taking $N\to\infty$ limit of those expressions yields
\begin{equation}
\lim_{N\to\infty}C_0=\begin{cases}
        1,\quad &J>-(\gamma+\omega_{\rm a})/2,\\
        0,\quad &\text{$N$ is even, $J<-(\gamma+\omega_{\rm a})/2$},\\
       \displaystyle \frac{\gamma}{\sqrt{\gamma^2+\omega_0^2}},\quad  &\text{$N$ is odd, $J<-(\gamma+\omega_{\rm a})/2$}.
    \end{cases}
\end{equation}

\begin{figure}
    \centering
    \includegraphics[width=1\linewidth]{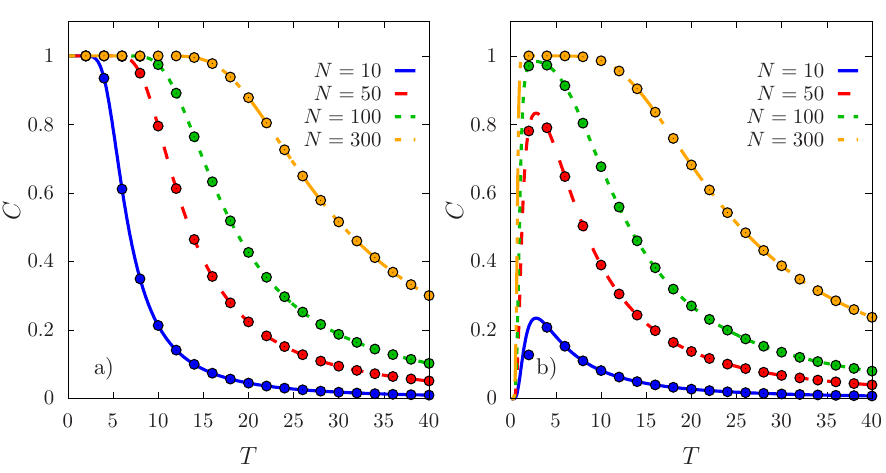}
    \caption{Comparing temperature dependence of exact $C$ [Eq.~\eqref{eq:C-R}, symbols] and its large-$N$ asymptotics [Eq.~\eqref{eq:C-largeN-new}, lines] for a) ferromagnetic and b) antiferromagnetic ground state. Parameters used are $\omega_0=1$, $\omega_{\rm a}=2$, $\gamma=3$, in a) $J=4$, and in b) $J=-4$. The figure demonstrates that once there are a few tens of source TLSs, a simple formula~\eqref{eq:C-largeN-new} captures the temperature dependence of $C(T)$ almost perfectly at $T>0$. With the increasing number of source TLSs, $C(T)$ gets closer to its maximum value of 1.}
    \label{fig:fig4_ClargeN}
\end{figure} 

On the ground-state crossing line~\eqref{eq:Jc-Neven} the coherence $C$ at $T=0$ is given by Eq.~\eqref{eq:C0-Jnegative-Neven-pt} for $N$ even. For $N$ odd, at the ground-state crossing line~\eqref{eq:Jcritodd}, we get 
\begin{equation}{\label{eq:C0-Jnegative-Nodd-pt}}
C_0=\frac{N}{N+1}\frac{\gamma}{\sqrt{\omega_0^2+\gamma^2}}+\frac{1}{N+1}\frac{\gamma N}{\sqrt{\omega_0^2+\gamma^2N^2}}.
\end{equation}
The $N\to\infty$ limit of Eqs.~\eqref{eq:C0-Jnegative-Neven-pt} and~\eqref{eq:C0-Jnegative-Nodd-pt} reads
\begin{equation}
\lim_{N\to\infty}C_0=\begin{cases}
1/3,\quad &\text{$N$ is even, $J=-(\gamma+\omega_{\rm a})/2$},\\
\displaystyle \frac{\gamma}{\sqrt{\gamma^2+\omega_0^2}},\quad  &\text{$N$ is odd, $J=-(\gamma+\omega_{\rm a})/2$}, \end{cases}
\end{equation}
which concludes the analysis of the large-$N$ limit of $C$ at $T=0$.

To derive the large-$N$ asymptotics of $C(T)$ at $T>0$, we start with the exact result~\eqref{eq:C-exact}. Since for any given $M>0$, the relative number of summands for which $|s|<M$ is $\simeq 2M/N\to 0$, the sums in Eq.~\eqref{eq:C-exact} are dominated by $|s|\sim N$, which represents the concentration of magnetization $2s$ of source TLSs around its most likely value. In Eq.~\eqref{eq:C-exact} we then can approximate $\sqrt{\omega_0^2+(2\gamma s)^2}\simeq 2\gamma|s|$, giving us
\begin{equation}
C \simeq \frac{Z_{\rm a}(h_+)-Z_{\rm a}(h_-)}{Z_{\rm a}(h_+)+Z_{\rm a}(h_-)},
\qquad h_\pm = \frac{\beta(\omega_{\rm a}\pm\gamma)}{2},
\end{equation}
where $Z_{\rm a}(h)=\sum_{s,d} \Omega(s,d,N)e^{2hs+K(N-4d)}$ is the partition function of the Ising chain (source TLSs) with dimensionless interaction $K=\beta J/2$ in the magnetic field $h$. By the transfer-matrix method~\cite{Folk/Holovatch:2024, Baxter:1982}, we get $Z_{\rm a}(h)=\lambda_+^N(h)+\lambda_-^N(h)$ with 
\begin{equation}
\lambda_\pm(h)=e^{K}\cosh(h)\pm\sqrt{e^{2K}\sinh^2(h)+e^{-2K}} .
\end{equation}
Neglecting $(\lambda_-/\lambda_+)^N\ll 1$, yields the approximation  
\begin{equation}
C(T) \sim \tanh\!\left[\frac{N}{2}
\ln\frac{\lambda_+(h_+)}{\lambda_+(h_-)}\right],\qquad N\to\infty, \qquad T>0. 
\label{eq:C-largeN-new}
\end{equation}
Since $\lambda_+(h)=\lambda_+(-h)$ is strictly increasing in $|h|$, and $|h_+|>|h_-|$ for $\gamma,\omega_{\rm a}>0$, the argument of $\tanh$ is positive. Hence the asymptotic approximation~\eqref{eq:C-largeN-new} is a monotonically increasing function of $N$, yielding $C\to 1$ as $N\to\infty$. Figure~\ref{fig:fig4_ClargeN} compares the exact $C(T)$ and its large-$N$ asymptotics for the ferromagnetic [panel a)] and antiferromagnetic [panel b)] ground states.

\section*{References}\vspace{-3ex}

%

\end{document}